\documentclass[12pt,preprint]{aastex}

\usepackage{graphicx}

\shorttitle{Effects of Reaction-Rate Variations on XRB Nucleosynthesis}
\shortauthors{Parikh et al.}

\newcommand{\msun}{M$_\odot$ }

\newcommand{\zapj}{ApJ~}

\newcommand{\gap}{\mathrel{ \rlap{\raise.5ex\hbox{$>$}}
      {\lower.5ex\hbox{$\sim$}} } }
\newcommand{\lap}{\mathrel{ \rlap{\raise.5ex\hbox{$<$}}
		 {\lower.5ex\hbox{$\sim$}} } }

\begin{document}

\title{The Effects of Variations in Nuclear Processes on 
       Type I X-Ray Burst Nucleosynthesis}

\author{Anuj Parikh\footnote{Current address: Physik Department E12,
       Technische Universit\"at M\"unchen, James-Franck-Strasse,
       D-85748 Garching, Germany; anuj.parikh@ph.tum.de}}
\affil{Departament de F\'\i sica i Enginyeria Nuclear, EUETIB,
      Universitat Polit\`ecnica de Catalunya, C./ Comte d'Urgell 187,
      E-08036 Barcelona, Spain; xrayburst@gmail.com}
\author{Jordi Jos\'e}
\affil{Departament de F\'\i sica i Enginyeria Nuclear, EUETIB,
      Universitat Polit\`ecnica de Catalunya, C./ Comte d'Urgell 187,
      E-08036 Barcelona,  and Institut d'Estudis Espacials de Catalunya (IEEC), 
      Ed. Nexus-201, C/ Gran Capit\`a 2-4, E-08034, Barcelona, Spain; 
      jordi.jose@upc.edu}
\author{Ferm\'\i n Moreno}
\affil{Departament de F\'\i sica i Enginyeria Nuclear, EUETIB,
      Universitat Polit\`ecnica de Catalunya, C./ Comte d'Urgell 187,
      E-08036 Barcelona, Spain; moreno@ieec.fcr.es}
\and
\author{Christian Iliadis}
\affil{Department of Physics and Astronomy, University of North Carolina,
       Chapel Hill, NC 27599-3255; and Triangle Universities Nuclear Laboratory,
       Durham, NC 27708-0308; iliadis@unc.edu}

\received{}
\accepted{}

\slugcomment{\underline{Submitted to}: \zapj \underline{Version}:
\today}

\clearpage

\begin{abstract}

 Type I X-ray bursts are violent stellar events that take place on the H/He-rich
 envelopes of accreting neutron stars.  We have investigated the role played by
 uncertainties in nuclear processes on the nucleosynthesis accompanying these explosive
 phenomena.
 Two different approaches have been adopted, in the framework of
 post-processing calculations. In the first one, nuclear
 rates are varied individually within uncertainties. 
 Ten different models,
 covering the characteristic parameter space for these stellar events,
 have been considered.
 The second, somewhat complementary approach involves a Monte Carlo
 code in which all nuclear rates are randomly varied within
 uncertainty limits  simultaneously. 
 All in all, about 50,000 post-processing calculations,
 with a network containing 606 nuclides (H to $^{113}$Xe) and
 more than 3500 nuclear processes, have been performed in this work.
 A brief comparison between both procedures is outlined together
 with an overall account of the key nuclear reactions whose
 uncertainties have the largest impact in our X-ray burst
nucleosynthesis studies.
\end{abstract}


\section{Introduction}

Type I X-ray bursts (hereafter, XRBs) are cataclysmic stellar events powered by 
thermonuclear runaways (TNR) in the H/He-rich envelopes 
accreted onto neutron stars in close binary systems (see reviews by
Bildsten 1998; Lewin et al.  1993, 1995; Psaltis 2006; Schatz \& Rehm 2006; 
Strohmayer \& Bildsten 2006).  They constitute the most frequent
type of thermonuclear stellar explosion in the Galaxy (the third, in terms of total energy output
after supernovae and classical novae) because of their 
short recurrence period (hours to days). More than $80$ Galactic low-mass X-ray 
binaries exhibiting such bursting behavior (with $\tau_{burst} \sim$ 10 - 100 sec) 
have been found since the discovery of XRBs by Grindlay et al. (1976), 
and independently, by Belian et al. (1976).

Modeling of type I X-ray bursts and their associated nucleosynthesis 
(see pioneering models by Woosley \& Taam 1976, Maraschi \& Cavaliere 1977, and Joss 1977) has been 
extensively addressed by different groups, reflecting the astrophysical interest in
determining the nuclear processes that power the explosion as well as
in providing reliable estimates for the chemical composition of the neutron star surface.
Indeed, several thermal (Miralda-Escud\'e, Paczy\'nski, \& Haensel 1990; Schatz et al. 1999), electrical 
(Brown \& Bildsten 1998; Schatz et al. 1999), and mechanical properties (Bildsten \& Cutler 1995;
Bildsten \& Cumming 1998) of the neutron star depend 
critically on the specific chemical abundance pattern of its outer layers.
The diversity of shapes in XRB light curves ($L_{peak} \sim 
3 \times 10^{38}$ erg s$^{-1}$ - see e.g. Galloway et al. (2007), Lewin et al. (1993), 
and Kuulkers et al. (2003)) is also likely due to different nuclear histories (see Heger et al. 2008,
for an account of the interplay between long bursts and the extension of the rp-process in XRBs),
suggesting that the final composition is not unique. 
It is worth noting that, as discussed by Taam (1980) and Woosley et al. (2004), the properties
of the bursts recurring in a given stellar source are affected by {\it compositional inertia};
that is, they are sensitive to the fact that accretion proceeds onto the ashes 
of previous bursts.  
Indeed, this compositional inertia seems to reduce
the expected recurrence times between bursts (particularly for scenarios involving accretion of
low-metallicity matter).
Moreover, these ashes may provide characteristic signatures such as gravitationally redshifted
atomic absorption lines from the surface of the neutron star, which could be identified
through high-resolution X-ray spectra (see Cottam, Paerels, \& Mendez 2002; 
Bildsten, Chang, \& Paerels 2003; Chang, Bildsten, \& Wasserman 2005; 
Chang et al. 2006; Weinberg, Bildsten, \& Schatz 2006), providing a valuable tool to constrain XRB models. 
Furthermore, the mechanism responsible for superbursts, highly energetic ($\sim10^{42}$ erg), 
long duration ($\tau \sim 10^4$ s) bursts (Cornelisse et al. 2000; Wijnands 2001, and references therein),
first reported from the source 4U 1735-44 (Cornelisse et al. 2000), also depends upon XRB nucleosynthesis.  
These phenomena are attributed to ignition in a C-layer accumulated from successive type I XRBs
(first proposed by Woosley \& Taam 1976; see also, Taam \& Picklum 1978, Brown \& Bildsten 1998,
Cumming \& Bildsten 2001, Schatz, Bildsten, \& Cumming 2003, Weinberg, Bildsten, \& Brown 2006,
and Weinberg \& Bildsten 2007).
The potential impact of XRB nucleosynthesis on Galactic
abundances is still a matter of debate: although ejection from a neutron star is unlikely because of 
its large gravitational potential (matter accreted onto a neutron star of mass $M$ and radius $R$ 
releases $G M m_p/R$ $\sim$ 200 MeV/nucleon, whereas only a few MeV/nucleon are released from thermonuclear fusion), 
radiation-driven winds during photospheric radius
expansion may lead to ejection of a tiny fraction of the envelope (containing nuclear
processed material --see Weinberg et al. 2006a and MacAlpine et al. 2007). Indeed, XRBs may help to explain the Galactic 
abundances of the problematic light {\it p-nuclei} (Schatz et al. 1998).

With a neutron star as the underlying compact 
object, temperatures and densities in the accreted envelope 
reach quite high values: $T_{peak} > 10^9$ K, and $\rho \sim 10^6$ g.cm$^{-3}$  
(note that during superbursts, however, densities may exceed $10^9$ g.cm$^{-3}$ --see Cumming \& Bildsten 2001). 
These values are at least an order 
of magnitude larger than in a typical classical nova outburst - see e.g. Jos\'e \& Hernanz (1998).  
As a result, detailed nucleosynthesis 
studies require the use of hundreds of isotopes,  up to the SnSbTe mass region
(Schatz et al. 2001) or beyond (the nuclear activity
in the XRB nucleosynthesis studies of Koike et al. (2004) reaches $^{126}$Xe),
and thousands of nuclear interactions. 
The main nuclear reaction flow is driven by the {\it rp-process} (rapid
proton-captures and $\beta^+$-decays), the $3\alpha$-reaction
and the {\it $\alpha$p-process} (a sequence of ($\alpha$,p) and (p,$\gamma$) reactions), and  
is expected to proceed far away from the valley of
stability, merging with the proton drip-line beyond A = 38 (Schatz et al. 1999).

Until recently, because of computational limitations, studies of XRB nucleosynthesis
have been performed using limited nuclear reaction networks, truncated 
around Ni (Woosley \& Weaver 1984; Taam et al. 1993; Taam, Woosley, \& Lamb 1996 --all using a 
19-isotope network), Kr (Hanawa, Sugimoto, \& Hashimoto 1983 --274-isotope network; 
Koike et al. 1999 --463 nuclides), Cd (Wallace \& Woosley 1984 --16-isotope network), 
or Y (Wallace \& Woosley 1981 --250-isotope network).  
On the other hand, Schatz et al. (1999, 2001)
have carried out very detailed nucleosynthesis calculations with a network containing 
more than 600 isotopes (up to Xe, in Schatz et al. 2001), 
but using a one-zone approach. Koike et al. (2004) have also performed detailed one-zone nucleosynthesis 
calculations, with temperature and density profiles obtained from a spherically symmetric evolutionary code,
linked to a 1270-isotope network extending up to $^{198}$Bi.

Note however that different numerical approaches and approximations 
(hydrodynamic simulations with limited networks or one-zone calculations with detailed networks) 
have been adopted in all those works, and hence, the predicted
nucleosynthesis in each case has to be taken with caution. 
Indeed, recent attempts to couple hydrodynamic stellar calculations (in 1-D)
and detailed networks include Fisker et al. (2004, 2006, 2007, 2008) and Tan et al. (2007) 
(using networks of $\sim$ 300 isotopes,
up to $^{107}$Te), Jos\'e \& Moreno (2006) (using a network of 2640 nuclear reactions and 478 isotopes, up to Te)
and Woosley et al. (2004) (using up to 1300 isotopes with an adaptive network).

Most of the nuclear reaction rates required for these extensive nucleosynthesis
calculations (in particular, those related to heavy species) rely on theoretical
estimates from statistical models, and therefore may be affected by significant
uncertainties. To date, only partial efforts have been made to quantify the impact of such nuclear
uncertainties on the overall abundance pattern accompanying XRBs 
(Wallace \& Woosley 1981; Schatz et al. 1998; Iliadis et al. 1999; Koike et al. 1999, 2004;
 Thielemann et al. 2001; Fisker et al. 2004, 2006, 2008; Amthor et al. 2006),
revealing a complex interplay between nuclear activity and the shape of the light curve
(Hanawa et al. 1983, Woosley et al. 2004). 
It has to be stressed, however, that simple, parametrized models cannot adequately address this issue,
since nuclear changes in the innermost envelope shell do not show up immediately at the surface
of the star. 

But in order to provide reliable estimates of the composition of neutron star envelopes 
it is of paramount importance to fully identify the key reactions whose uncertainties
have the largest impact on XRB yields (reactions that probably deserve further
improvement through nuclear physics experiments at dedicated facilities). 
To achieve such a goal, we have performed a comprehensive study of 
the effects of thermonuclear reaction-rate variations on type I X-ray burst nucleosynthesis,
sampling the overall parameter space. The scale of this study makes 
state-of-the-art hydrodynamic models computationally prohibitive.
Hence, we rely on post-processing calculations, coupling 
a detailed and fully updated nuclear reaction network
with temperature and density profiles extracted from literature. 

   The structure of the paper is as follows.  In Section 2, we outline the  
two approaches adopted in this paper, in the framework
of post-processing calculations with temperature and density profiles extracted
from the literature.  In Section 3, we extensively describe the results obtained from
the different models, through a sensitivity study in which all reaction rates 
are varied individually within uncertainty limits.
This is compared with Monte Carlo simulations in which all nuclear
reactions are simultaneously varied with randomly sampled enhancement factors (Section 4).
A summary of the main conclusions reached in this study, including
a list of key reactions whose uncertainties deeply influence the final type I XRB yields,
is presented in Section 5.

\section{Models \& Input Physics}

\subsection{Rate uncertainties and post-processing calculations}
 Nucleosynthesis studies in XRB conditions are computationally challenging
 because of the characteristic timescales and peak temperatures achieved,
 the large number of nuclear species involved (several hundred), and 
 the huge networks (with several thousand nuclear processes)
 required to address the rp- and $\alpha$p-processes occuring for H- and He-rich
 mixtures accreted onto the surface of a neutron star. 
 
 Two different, somewhat complementary approaches, based on 
 post-processing calculations with temperature and density profiles, can be adopted in a 
 comprehensive sensitivity study. 
 In the first one, all rates are varied individually (see Section 3) within 
 uncertainty limits so as to check the impact of each nuclear process
 on the final yields.  At least 2 post-processing calculations are required per
 nuclear rate to account for the upper and lower limits posed by its associated 
 uncertainty. For a network containing
 several thousand interactions the overall number of post-processing calculations is
 indeed extraordinarily large.
 It is worth mentioning that this technique has been previously applied to a large number of astrophysical sites,
 including nucleosynthesis in the Sun (Bahcall et al. 1982), type II supernovae
 (The et al. 1998; Jordan, Gupta, \& Meyer 2003), classical nova explosions
 (Iliadis et al. 2002), primordial (Big Bang) nucleosynthesis (Coc et al. 2002, 2004), 
 intermediate-mass AGB stars (Izzard et al. 2007), and Type I X-ray bursts (Amthor et al. 2006).
 
 A second approach is based on Monte Carlo techniques (see Section 4). Here, random enhancement factors
 (often adopted to follow a log-normal  distribution) are applied
 to each nuclear process of the network simultaneously. The impact on the
 final yields is then tested through a series of post-processing calculations. This approach
 requires a large number of trial simulations in order to be statistically sound
 and has been already applied to Big Bang nucleosynthesis studies (Krauss \& Romanelli 1990;
 Smith et al. 1993), nova nucleosynthesis (Smith et al. 2002; Hix et al. 2002, 2003) and
 also to Type I X-ray bursts (Roberts et al. 2006).

 Both approaches have been adopted in this study (although more emphasis has been placed
 on the first one).  An interesting issue, recently raised by Roberts et al. (2006), is the feasibility
 of the first method, as compared with the Monte Carlo approach, to properly address
 the higher-order correlations between input rates and XRB model predictions because of the
 large number of reactions simultaneously involved in the production and destruction of each element.
  It is also the goal of this manuscript to test this conjecture
 by comparing the results from an individual-variation sensitivity study with 
 those obtained with a Monte Carlo approach.

\subsection{Initial Models: Temperature and Density profiles}
 
 In this work, we have used three temperature-density versus time profiles directly
 extracted from the literature (see Fig. 1).  Model K04 (Koike et al. 2004) is based on a spherically
 symmetric model of accretion onto a 1.3 \msun neutron star (with R$_{\rm NS}$=8.1 km).
 This model is characterized by a
 peak temperature of 1.36 GK, densities ranging between (0.54 - 1.44) $\times
 10^6$ g.cm$^{-3}$, and a burst duration of $\sim 100$ s. The initial composition
 is taken directly from Koike et al. (2004), with X=0.73, Y=0.25 and Z=0.02 --roughly solar.

Model F08 (Fisker et al. 2008) is based on 1-D general relativistic hydrodynamic calculations. 
It achieves a peak temperature of only 0.993 GK, densities ranging 
between (2.07 - 5.14) $\times
10^5$ g.cm$^{-3}$, and a burst duration of $\sim 50$ s. The initial composition
is taken directly from the hottest burning zone of the
hydrodynamic models of Fisker et al. (2008), with X=0.40, Y=0.41 and Z=0.19 --a metallicity about ten times solar; 
see the original manuscript for a detailed account of
the distribution of those metals.

Finally, Model S01 (based on Schatz et al. 2001) achieves the largest temperature of all models 
extracted from the literature, with $T_{peak} = 1.907$ GK. 
The shape of this temperature profile, based on one-zone model calculations, is quite different
from those corresponding to Models K04 and F08 (which approximately follow an exponential decay): 
indeed, it shows a long plateau at about T $\sim$1.4 GK before the final decline. 
Since no density
profile is directly available for this model, and considering the relatively small variation in density during an XRB,
we have scaled K04 to match the values reported in Schatz et al. (2001),
resulting in densities ranging between (0.54 - 1.73) $\times 10^6$ g.cm$^{-3}$. The  
duration is $\sim 300$ s. 
The initial composition
is X=0.718, Y=0.281 and Z=0.001 --a metallicity about 20 times lower than solar. 
Because of the lack of information on the specific distribution of the metallicity, we assume
that all metals correspond to $^{14}$N (see also Woosley et al. 2004), following the rapid rearrangement of
CNO isotopes that naturally occurs early in the burst.

 The three models mentioned above partially cover
 the parameter space in XRB nucleosynthesis calculations. However, it is difficult,
 if not impossible, to disentangle the specific role played by the initial metallicity, the
 peak temperature achieved and the duration of the bursts, since all these models are characterized by
 different values for these physical quantities.
  Hence, to appropriately address this issue, we have generated additional models through parameterization 
of the  Koike et al. (2004) temperature-density-time profiles. 
Parameterized models have been used previously to study, for example, nucleosynthesis in novae (Boffin et al. 1993), 
the alpha-rich freeze-out in Type II supernovae (The et al. 1998, Jordan et al. 2003) and XRBs (Wallace and Woosley 1981, 1984).

To evaluate the role played by the duration of the burst (taken as the characteristic timescale of the temperature and
density profiles) in the extension of the 
nuclear path, we have scaled the K04 profiles in duration by a factor 0.1 (short burst: Model K04-B1) and 10 
(long burst: Model K04-B2) while preserving Tpeak and the initial chemical composition.

The role played by the initial metallicity (reflecting that of the stream of infalling
material during accretion) has been tested through 3 models: a low-metallicity
model (Z=$10^{-4}$; Model K04-B3), and two high metallicity ones (with Z=0.19; Models K04-B4 \&
K04-B5).  The initial composition of Models K04-B3 and K04-B4 is scaled from the
distribution reported in Koike et al. (2004). In contrast, for Model K04-B5, we have adopted
the distribution given in Fisker et al. (2008). This will also allow us to test
if a different distribution of metals has an impact on the final yields.  

Finally, two additional models have been constructed to test the effect of varying the peak temperature
achieved during the explosion: we have scaled the temperature and density versus time profiles from Koike et al. (2004) 
to attain peak values of $0.9 \times 10^9$ K (lower peak temperature; Model K04-B6) 
and $2.5 \times 10^9$ K (higher peak temperature; Model K04-B7), while preserving
the initial composition and the duration of the burst.

 It is worth noting that post-processing calculations
 are not suited to derive absolute abundances (or, as mentioned above, to provide any insight into
 light-curve variations and energetics) since they rely only on 
 temperature and density versus time profiles evaluated at a given location of the star
 (usually, the innermost shells of the envelope). Indeed, it is likely that 
 the evolution at other depths will be characterized by a different set of 
 physical conditions. Furthermore, adjacent shells will eventually mix when 
 convection sets in, altering the chemical abundance pattern in those layers.
 However, this approach is reliable to identify the key processes governing
 the main nuclear activity at the specific temperature and density regimes
 that characterize such bursting episodes. 
 Hence, the goal of this paper is to provide the reader with a list of key
 nuclear processes whose uncertainties have the largest influence on the final yields,
 covering as much as possible the proper range of temperatures, densities, and timescales
 that characterize XRBs.  Any attempt to properly
 quantify the extent of this impact (for instance, in terms of absolute
 abundances) however, must rely on state-of-the-art hydrodynamic codes coupled to detailed
 nuclear networks.
 
\subsection{Nuclear Reaction Network}

The temperatures and densities achieved in XRB nucleosynthesis are sufficiently high so that many 
nuclear reactions, especially those with relatively small Q-values, achieve an equilibrium between 
the forward and reverse process. For example, if the reaction $A(p,\gamma)B$ has a small Q-value, then the strong 
$B(\gamma,p)A$ photodisintegration will give rise to a small equilibrium abundance of nuclei $B$ that may then 
capture another proton. Such cases, which are called (sequential) two-proton captures, must be considered 
carefully since they represent waiting points (and may even be candidates for {\it termination points}) for 
a continuous abundance flow toward heavier-mass nuclei. Among the most important of these waiting points are 
$^{64}$Ge, $^{68}$Se and $^{72}$Kr. When a reaction rate equilibrium has been established, the most 
important nuclear physics information needed is the reaction Q-value (which enters exponentially in the 
effective decay constant of $A$) of the link $A(p,\gamma)B$ and the reaction rate of the second-step 
reaction $B(p,\gamma)C$, but the {\it rate} of the reaction $A(p,\gamma)B$ is irrelevant. Note that for 
the three nuclei mentioned above, the (p, $\gamma$) Q-values have not been measured directly yet. According to Audi et al. (2003a), 
their predicted values amount to $Q= - 80\pm300$ keV, $- 450\pm100$ keV and $- 600\pm150$ keV, respectively.
Some encouraging progress has been made through
mass measurements of $^{64}$Ge (Clark et al. 2007, Schury et al. 2007), $^{68}$Se (Clark et al. 2004,
Wohr et al. 2004, Chartier et al. 2005), and $^{72}$Kr (Rodr\'\i guez et al. 2004). 

In order to set up the reaction library for our study, we started by adopting the proton drip line from 
Audi et al. (2003a,b). For reactions $A(p,\gamma)B$ with Q-values below 1 MeV the rates (below the element Pd) are 
calculated using the Hauser-Feshbach code MOST (Goriely 1998; Arnould \& Goriely 2006) and the corresponding reverse 
photodisintegrations are computed with the {\it extrapolated} proton separation energies from Audi et al. (2003a). 
Notice that nuclear masses enter twice:
first in the calculation of the forward Hauser-Feshbach rate, and second in the calculation of the reverse 
photodisintegration rate. The second-step reactions, $B(p,\gamma)C$, are also computed with the code MOST. We had to 
include many nuclides beyond the proton drip line in order to properly account for sequential two-proton captures. 
For all other reactions for which experimental rates are not available, we used the results from the Hauser-Feshbach 
code NON-SMOKER (Rauscher \& Thielemann 2000).

Our entire nuclear network consists of 606 nuclides ranging from $^{1}$H to $^{113}$Xe. All charged-particle 
induced reactions between these isotopes, and their corresponding reverse processes, have been included. 
Experimental information was only available for the rates of a small subset of reactions.
These rates are adopted from Angulo et al. (1999), 
Iliadis et al. (2001), and some recent updates for selected reactions. Neutron captures are disregarded since our 
early test calculations have revealed that they play a minor role in XRB nucleosynthesis. All reaction rates incorporate
the effects of thermal excitations in the target nuclei (Rauscher \& Thielemann 2000).

For the weak interactions, we use laboratory decay rates (Audi et al. 2003b); the impact 
of $\beta$-delayed nucleon emission has also been considered. For a discussion of employing {\it stellar} 
versus {\it laboratory} decay rates, see Woosley et al. (2004). Also, note that many computed stellar 
decay rates (Fuller, Fowler \& Newman 1982a,b) do not converge to their laboratory values at lower 
temperatures and densities. In order to avoid this problem we decided against using stellar decay rates. Although 
the use of properly converging stellar decay rates is in principle preferred, we doubt that the effect 
would be large in XRB nucleosynthesis calculations.

\section{Individual Reaction-Rate Variations}

Some general procedures must be outlined before a detailed analysis of
the main results is presented. 
For all 10 models considered in this Section, 
a first calculation, with {\it standard} rates (as described
in Section 2.3), has been performed.
This is used to scale the level of changes (indicated in Tables 1--10, \& 13--14 as 
$X_i/X_{i,std}$, that is, the ratio of mass fractions obtained with a modified
network -resulting from our exploration of uncertainties- to
those obtained with our {\it standard} network).
Following this, a whole series of post-processing calculations have been computed, in which each
nuclear rate is varied individually by a factor 10 up and down,
for each of the 10 models.

Several important remarks have to be made at this stage: first, 
we have chosen a factor of 10 
for the level of uncertainty affecting theoretical reaction rate estimates, in general. 
Other authors (see Schatz 2006, and Amthor et al. 2006) claim, instead, that excitation
energies of theoretically calculated levels for XRB conditions may suffer 
uncertainties of $\sim$ 100 keV, which translates into an overall uncertainty in some 
rates that may reach several orders of magnitude. It is also worth noting that
efforts to ascertain systematic uncertainties through a direct comparison between
different theoretical models (see Section 5.1) often report variations smaller
than a factor of 10.
Second, some nuclear rates, such as the triple-$alpha$ reaction, some (p, $\gamma$) reaction rates on low-mass targets, or
most of the $\beta$-decay rates (note again that we used {\it ground-state} or {\it laboratory} $\beta$-decay rates), 
are known with much better precision. This has 
been addressed when necessary and is discussed in the following sections of this manuscript.
Third, it is important to stress that forward and reverse reactions have been 
varied simultaneously with the same uncertainty factor, to preserve the principle of detailed balance.
Fourth, a particular strategy has been adopted to evaluate the impact of 
$\beta$-decay rate uncertainties: they have initially been varied by a factor of 10,
up and down as a way to discriminate the {\it key} $\beta$-decay rates
from the rest (see Table 11). Then, additional post-processing calculations focused on these key rates
have been performed assuming more realistic uncertainties (around 30\%, according
to ground-state $\beta$-decay rate uncertainties; see Audi et al. 2003b).
Fifth, some additional  tests aimed at identifying the influence of
Q-value variations have been performed using the uncertainties
estimated by Audi et al. (2003a) (see Tables 12--14). This part of our study has been restricted to
Models K04 and F08, to proton-capture reactions on  
$A < 80$ nuclei, and to reactions with $\left|Q\right|<$ 1 MeV, for which the
estimated uncertainty exceeds 50 keV (see Table 12).   
Sixth, since the main goal of this paper is to identify 
the key nuclear processes whose uncertainties have the largest impact on XRB 
nucleosynthesis, we have restricted
the analysis (in the Tables and forthcoming discussion) to nuclear species which achieve a mass fraction of
at least $10^{-5}$ at the end of the burst, and deviate from the abundances computed
with standard rates by at least a factor of 2. 
And seventh, for the sake of clarity and conciseness, 
the isotopes displayed and discussed throughout this manuscript correspond
 to species that are either stable or
have a half-life longer than 1 hour (the rest are assumed to fully decay 
at the end of the burst and 
consequently, are added to the corresponding stable or long-lived daugther nuclei). 

\subsection{Results}

In this Section, we report  results from a series of $\sim$40,000 post-processing calculations 
(requiring 14 CPU-months) performed for this study of X-ray burst
nucleosynthesis.

\subsubsection{Model K04}

Here, we will describe in detail the results obtained from our post-processing
calculations using temperature and density versus time profiles extracted
from Koike et al. (2004) [Model K04]. A general discussion of the main nuclear path
achieved for these profiles can be found in Iliadis (2007).  
Our main results, from individually varying each reaction rate by a factor 10 (up and down), are summarized in Table 1.

Because of the moderate peak temperature achieved, the extent of the nuclear
activity (defined as the heaviest isotope with $X_i > 10^{-2}$, in the final yields) 
reaches $^{96}$Ru. The most abundant species (stable or with a half-life $>$ 1 hr) 
at the end of the burst are H (0.20, by mass), $^4$He (0.021), 
$^{68}$Ge (0.20), $^{72}$Se (0.13), $^{64}$Zn (0.071), and $^{76}$Kr (0.074).
Qualitatively, this is in agreement with the results reported from Koike
et al. (2004) (Model 2, Table 8), in which the most abundant 
species,  at the
end of the burst, are also $^{68}$Ge (0.17), $^{72}$Se (0.14), $^{64}$Zn (0.27), 
and $^{76}$Kr (0.078).
Note that the final mass fraction of
$^4$He reported by Koike et al. (2004), 0.011, is also comparable to the value reported in this paper. However,
Koike's model is fully depleted of H ($9.76 \times 10^{-17}$, by mass), whereas
some H remains at the end of our simulations. This could result from differences
in the adopted nuclear reaction 
networks as well as to the hybrid use of evolutionary and post-processing calculations in Koike et al. (2004).
 This discrepancy is of some importance as the presence of H in the ashes may induce
marginal nuclear activity driving a second burst if the set of necessary conditions
for a TNR are satisfied (see Woosley \& Weaver 1984, for details).

Our study reveals that the most influential reaction, by far, is the 3 $\alpha$,
which affects a large number of species ($^4$He, $^{18}$F, $^{21}$Ne, $^{24,25}$Mg,
$^{28-30}$Si, $^{33,34}$S,  $^{36-38}$Ar, $^{41}$Ca, $^{44,46,47}$Ti, $^{49}$V,  
 $^{50}$Cr, $^{53}$Mn, $^{54}$Fe, $^{57-59}$Ni, $^{61,63}$Cu, $^{62}$Zn, $^{102}$Pd,
and $^{103,104}$Ag), when its nominal rate is varied by  a factor of 10, up and down.
A similar result on the importance of the 3 $\alpha$ reaction has been previously 
reported by Roberts et al. (2006), using a Monte Carlo approach.
But, as discussed above, this uncertainty factor is far too large and accordingly, this reaction
 has been removed from Tables 1-10. Hereafter, we will drop from the discussion of the different models any additional
comment on the importance of the 3$\alpha$ reaction which, when arbitrarily varied by a factor of
10, systematically becomes the single most influential reaction of the whole network, for all 10 models.
Indeed, additional tests performed with a more realistic uncertainty\footnote{Note that 
according to Tur et al. (2006) the 3$\alpha$ rate is known to $\pm$ 12\%.} of $\pm$40\%
(see Angulo et al. 1999) show no effect on any individual isotope, for any of the 10 models 
(the impact of nuclear uncertainties affecting the triple alpha rate on the total energy output
will be specifically addressed in Section 5). 

From the several thousand nuclear processes, we find only 56 reactions--and the corresponding reverse 
reactions-- have an impact on the
final yields when their rates are varied by a factor of 10, up and down for this model (see Table 1).
Furthermore, our study reveals that the impact of most of these reactions
is restricted to the vicinity of the target
nuclei. A clear example is $^{15}$O($\alpha$,$\gamma$): when its nominal
rate is multiplied by a factor 0.1, only one isotope, $^{15}$N,
is modified by more than a factor of 2 (as compared with the mass fraction
obtained with the recommended rate --but see also Section 5). Indeed, the most influential reaction is by
far $^{65}$As(p,$\gamma$), perhaps expected due to its bridging effect
on the  $^{64}$Ge-waiting point. To a lesser extent, $^{96}$Ag(p,$\gamma$), 
and  $^{102}$In(p,$\gamma$) also show an impact on a number of nuclear species. 

The most important $\beta$-decay rates identified in this Model are those of
$^{64}$Ge, $^{68}$Se, $^{72}$Kr, $^{76}$Sr, $^{80}$Zr, $^{88}$Ru,
$^{92}$Pd, and $^{99}$In (see Table 11). But it is important to stress that
additional calculations, in which these (laboratory) $\beta$-decay rates 
were varied within realistic uncertainties (half-lives varied by $\sim$ $\pm$30\%), have revealed
no effect on any single isotope  
(see Woosley et al. 2004, for a sensitivity study based on variations of groups
of weak rates, including all positron emission rates for nuclei heavier than $^{56}$Ni,
by an order of magnitude).

From the list of nuclear reactions selected for the Q-value variation study (see Table 12, and the
discussion in Section 3), only
 $^{26}$P(p, $\gamma$)$^{27}$S, $^{46}$Cr(p, $\gamma$)$^{47}$Mn,
$^{55}$Ni(p, $\gamma$)$^{56}$Cu, $^{60}$Zn(p, $\gamma$)$^{61}$Ga, 
and $^{64}$Ge(p, $\gamma$)$^{65}$As show some impact
on the final yields when their Q-values are varied between one sigma uncertainty bounds.
However, as summarized in Table 13, the effects are restricted to single 
isotopes, except for the last reaction, $^{64}$Ge(p, $\gamma$), whose influence ranges 
between $^{64}$Zn and $^{104}$Ag.
Indeed, the significance of this reaction on the $^{64}$Ge waiting point is well-known 
(see e.g., Woosley et al. 2004, Thielemann et al. 2001, Schatz 2006, 
Fisker et al. 2008, and references therein). 
Mass measurements on $^{64}$Ge have indeed been performed (Clark et al. 2007; Schury et al. 2007).
When combining these measurements with the $^{65}$As mass value given by Audi et al. (2003a),
the corresponding one sigma uncertainty limits in the $^{64}$Ge(p, $\gamma$)$^{65}$As Q-value 
($-74 \pm 303$ keV, and $-46 \pm 302$ keV, respectively) 
are well covered by the range used in our study (Q = $-80 \pm 300$ keV;  Audi et al. (2003a)).  Clearly,
mass measurements on $^{65}$As are still required to better determine this
Q-value.  These are challenging experiments because of the difficulty
of producing $^{65}$As (Clark, private communication).

\subsubsection{Model S01}

Here, we summarize the main results obtained from our post-processing
calculations with the temperature versus time profile extracted
from the one-zone model of Schatz et al. (2001) [Model S01]. 

Because of the larger  peak temperature achieved in this model, the main nuclear activity 
extends to heavier nuclei ($^{107}$Cd, for $X > 10^{-2}$) than in Model K04. 
The most abundant species (stable or with a half-life $>$ 1 hr) 
at the end of the burst are now H (0.071, by mass), $^4$He (0.013), 
and the heavy isotopes $^{104}$Ag (0.328), $^{106}$Cd (0.244), 
$^{103}$Ag (0.078), and $^{105}$Ag (0.085). This chemical pattern bears some resemblance 
to that reported by Schatz et al. (2001), and also to a similar model (zM) computed
by Woosley et al. (2004) with a 1-D hydrodynamic code.  For example, a large overproduction of 
the p-nucleus $^{106}$Cd is explicitly mentioned in Schatz et al. (2001), Fig. 4; as well,
Woosley et al. (2004) (Fig. 7) find a large mass fraction of $^{106}$Sn 
(which will decay into $^{106}$Cd), followed by $^{64}$Ge, $^{68}$Se, $^{104}$In 
(which will decay into $^{104}$Ag), $^{4}$He, and H. It is worth noting that the final amount of H is much
smaller than in Model K04: here, H and He are almost
depleted and hence the next burst will necessarily require the piling-up of fresh H/He-rich
fuel on top of this H/He-depleted shell.

Only 64 reactions (and the corresponding reverse
processes) from the overall nuclear network turn out to have an impact on the final 
yields when their rates are varied by a factor of 10, up and down (Table 2). 
As reported for Model K04, the influence of these reactions is often limited to the vicinity of the target nuclei.
Remarkable exceptions are $^{56}$Ni($\alpha$,p), $^{59}$Cu(p,$\gamma$),
 and to a lesser extent, $^{103}$Sn($\alpha$, p). Notice also that, among the most important reactions, 
only three $\alpha$-captures ($^{15}$O($\alpha$,$\gamma$), 
$^{56}$Ni($\alpha$,p), and $^{103}$Sn($\alpha$, p)) are found.
The role played by $^{15}$O($\alpha$,$\gamma$) is as marginal as for Model K04 however (but see discussion in
Section 5).

Concerning the most influential $\beta$-decay rates, 
we found similarities with the results reported for Model K04. Hence,
the $\beta$-decay rates of $^{68}$Se, $^{72}$Kr, $^{80}$Zr, $^{88}$Ru, and $^{92}$Pd,
are important, but due to the higher peak temperatures reached in this Model,
 the list also extends to heavier species, including  
$^{101-106}$Sn and $^{106}$Sb (see Table 11). Additional studies of the specific impact 
of these decays with realistic uncertainty bounds (Audi et al. 2003b) have also been
performed for this case, revealing no effect on any particular isotope. 

\subsubsection{Model F08}

Since this model achieves the lowest peak temperature of the three models
extracted from the literature, its main nuclear path is somewhat more
limited, reaching only (X $> 10^{-2}$) $^{72}$Se:
the most abundant species (stable or with a half-life $>$ 1 hr) 
at the end of the burst are now $^4$He (0.085), 
$^{56}$Ni (0.13), $^{60}$Ni (0.38), $^{12}$C (0.040), $^{28}$Si (0.041), and $^{64}$Zn (0.034).
Moreover, H has been depleted down to $10^{-11}$ by mass.  Qualitatively, there is 
good agreement with the relevant
chemical pattern reported in Fisker et al. (2008), Fig. 20. It is worth noting that the
large amount of  $^{12}$C obtained at the end of the burst may have implications for 
the energy source that powers superbursts (see Sect. 1). 

Again, only a few reactions --62 and the corresponding reverse reactions-- have an impact on the
final yields when their rates are varied by a factor of 10, up and down (Table 3). 
The most influential reactions are $^{12}$C($\alpha$,$\gamma$), $^{26g}$Al($\alpha$, p),
$^{57}$Cu(p,$\gamma$), $^{61}$Ga(p,$\gamma$), and $^{85,86}$Mo(p,$\gamma$). Additional
discussion and tests on the impact of the $^{12}$C($\alpha$,$\gamma$) rate are presented
in Section 5.
Thrirteen $alpha$-capture reactions show some impact
on the yields, a larger number than in K04 and S01.
Indeed, all key reactions affecting $A < 30$ species are mainly ($\alpha$,$\gamma$)
or ($\alpha$, p) reactions. It is worth mentioning, however, that $^{15}$O($\alpha$,$\gamma$) 
is absent from this list (see Sect. 5).

The most influential $\beta$-decay rates  
are now shifted towards lighter unstable isotopes, because of the lower $T_{peak}$ value achieved. These
include decays of $^{26,29}$Si, $^{29,30}$S,
$^{33,34}$Ar, $^{37,38}$Ca, $^{60}$Zn, $^{68}$Se, $^{72}$Kr, $^{76}$Sr, and $^{80}$Zr (Table 11).  
As in K04 and S01 though, no effect is seen on any isotope when realistic uncertainties are adopted
for these decay rates.

Finally, the only nuclear reaction from Table 12 that has a significant impact 
on the final yields when its Q-value is varied between one sigma uncertainty bounds is $^{45}$Cr(p, $\gamma$) 
(see Table 14). 

\subsubsection{Effect of the duration of the burst: Model K04-B1 vs. K04-B2}

Because of the shorter duration of Model K04-B1, as compared with Model K04,
 the main nuclear path extends up to $^{68}$Ge.
In contrast, the longer burst of Model K04-B2 drives the main path 
up to $^{106}$Cd. 

The most abundant species at the end of the bursts are:  H (0.42, by mass), $^4$He (0.061), 
$^{60}$Ni (0.10), $^{64}$Zn (0.31), and $^{68}$Ge (0.023), for
the shorter burst of Model K04-B1; and 
$^4$He (0.0051), $^{68}$Ge (0.159), $^{72}$Se (0.121), $^{76}$Kr (0.077), 
$^{80}$Sr (0.040), $^{90}$Mo (0.042), $^{94}$Mo (0.055), $^{98}$Ru (0.040), 
$^{102}$Pd (0.034), $^{103}$Ag (0.060), $^{104}$Ag (0.097), and $^{105}$Ag (0.042), 
for Model K04-B2. The final amount of H in the latter model has decreased down to
$10^{-15}$. The larger extension of the main nuclear activity reported for Model K04-B2 
is a direct consequence of the longer exposure times to higher temperatures. However,
the depletion of H results from the complex interplay between exposure time to high
temperatures, which decreases the overall H content, and the effects of photodisintegrations,
which raise the H abundance.

Concerning nuclear uncertainties, Model K04-B1 is characterized by 28 critical
reactions, the most important ones being 
$^{61}$Ga(p,$\gamma$) and $^{65}$As(p,$\gamma$),
with a marginal role played by $^{18}$Ne($\alpha$,p) and  $^{31}$Cl(p,$\gamma$)
(Table 4).  For Model K04-B2, with 51 critical reactions, the most important are  $^{65}$As(p,$\gamma$) and 
 $^{32}$S($\alpha$,$\gamma$), followed, to some extent, by 
 $^{12}$C($\alpha$, $\gamma$), $^{61}$Ga(p, $\gamma$), 
$^{75}$Rb(p, $\gamma$), $^{84}$Zr(p, $\gamma$), $^{92}$Ru(p, $\gamma$), $^{93}$Rh(p, $\gamma$),
and $^{96}$Ag(p, $\gamma$) (Table 5).
 Only 2 $\alpha-capture$ reactions appear to
 be influential in Model K04-B1 (one being $^{15}$O($\alpha$,$\gamma$), see Table 4), whereas
 uncertainties affecting 12 $\alpha$-capture reactions turn out to be critical
 for Model K04-B2 (see Table 5). 

Table 11 lists the most important $\beta$-decay rates for these models, when these 
rates are varied by a factor of 10 up and down.  In summary, $\beta$-decay rates of 
$^{25}$Si, $^{33}$Ar, $^{36-38}$Ca, $^{41}$Ti, $^{59}$Zn, $^{62-64}$Ge,
and $^{68}$Se, become critical in Model K04-B1, whereas those of $^{25}$Si, $^{68}$Se, $^{72}$Kr, $^{76}$Sr, $^{80}$Zr, $^{92,94}$Pd, and
$^{101-104}$Sn play a major role in Model K04-B2. As before, no noticeable effect remains when
realistic uncertainties are adopted for these rates.

\subsubsection{Effect of the initial metallicity: Models K04-B3, K04-B4, and K04-B5}

In the low metallicity Model K04-B3, the main nuclear path ($X > 10^{-2}$) stops at $^{96}$Ru.  
In contrast, the higher metallicity models K04-B4 and K04-B5 reach $^{68}$Ge and $^{72}$Se, respectively. 
The most abundant species at the end of the bursts are:  
    H (0.194), $^4$He (0.021), $^{68}$Ge (0.205), $^{72}$Se (0.132), $^{64}$Zn (0.075), 
$^{76}$Kr (0.073), $^{80}$Sr (0.041), and $^{82}$Sr (0.023), for Model K04-B3;
   $^4$He (0.018), $^{60}$Ni (0.696), $^{64}$Zn (0.161), $^{56}$Ni (0.051),
   $^{32}$S (0.014), and $^{68}$Ge (0.016), for Model K04-B4; and finally,
$^4$He (0.018), $^{56}$Ni (0.273), $^{60}$Ni (0.256), $^{39}$K (0.057), $^{64}$Zn (0.062), 
   and $^{68}$Ge (0.041), for Model K04-B5. Note that H has been seriously depleted in both
   high metallicity models, achieving mass fractions of  $4 \times 10^{-14}$ and  $10^{-15}$,
   respectively.

Concerning nuclear uncertainties, Model K04-B3 is characterized by 56 critical
reactions, while Models K04-B4 and K04-B5, have 43 and 45, respectively. 
The most important ones for Model K04-B3 (Table 6) are 
$^{65}$As(p,$\gamma$) and $^{96}$Ag(p,$\gamma$), with a
marginal role played by $^{102,103}$In(p,$\gamma$). 
For Model K04-B4 (Table 7), the most influential reactions are
$^{12}$C($\alpha$,$\gamma$), 
$^{30}$S($\alpha$,p), and to some extent,  $^{30}$P($\alpha$,p), and $^{65}$As(p,$\gamma$).  
Finally, for Model K04-B5 (Table 8), the key reactions are 
 $^{30}$S($\alpha$,p), $^{25}$Si($\alpha$,p), $^{59}$Cu(p,$\gamma$), 
$^{56}$Ni($\alpha$,p), $^{29}$S($\alpha$,p), $^{65}$As(p,$\gamma$), and $^{12}$C($\alpha$,$\gamma$), 
and to some extent, $^{61}$Ga(p,$\gamma$).
 Note that $^{15}$O($\alpha$,$\gamma$) is only important for Model K04-B3.
 The number of influential $\alpha$-induced reactions seems to increase with metallicity: only 2
reactions of this type are important in Models K04 and  K04-B3, whereas 22 and 17 
are important for Models K04-B4 and K04-B5, respectively.

Finally, the most influential $\beta$-decays from Model K04-B3 are $^{68}$Se,
$^{72}$Kr, $^{76}$Sr, $^{80}$Zr, $^{88}$Ru, $^{92}$Pd, and $^{99}$In.
Model K04-B4 is characterized by the importance of the beta-decays of
$^{33,34}$Ar, $^{37-39}$Ca, $^{42}$Ti, $^{46}$Cr, $^{49}$Fe, $^{55}$Ni,
$^{58}$Zn, $^{68}$Se, $^{72}$Kr, $^{76}$Sr, and $^{80}$Zr,
whereas for Model K04-B5, the beta-decay rates of 
$^{21}$Mg, $^{24,25}$Si, $^{28-30}$S, $^{33,34}$Ar,
$^{37,38}$Ca, $^{64}$Ge, and $^{68}$Se, 
are the most important (see Table 11). 
Again, no effect shows up when realistic
(ground-state) uncertainties are adopted for these rates.

\subsubsection{Effect of the peak temperature: Model K04-B6 vs. K04-B7}

In Model K04, the main nuclear path ($X > 10^{-2}$) reached $^{96}$Ru.
Because of the lower peak temperature achieved in Model K04-B6,
 the main nuclear path reaches $^{82}$Sr.
In contrast, the higher temperatures achieved in Model K04-B7 drives the main path
up to $^{103}$Ag. 

The most abundant species at the end of the burst are now  
H (0.151), $^4$He (0.034), $^{64}$Zn (0.375), $^{68}$Ge (0.193), $^{60}$Ni (0.051), 
$^{72}$Se (0.074), $^{76}$Kr (0.031), $^{80}$Sr (0.015), and $^{82}$Sr (0.011), for
the lower peak temperature Model K04-B6 (Table 9); and  
H (0.460), $^4$He (0.013), $^{68}$Ge (0.058), $^{72}$Se (0.069), $^{76}$Kr (0.048), 
$^{80}$Sr (0.031), and $^{96}$Ru (0.026), for Model K04-B7 (Table 10). It is worth noting
that Model K04-B7 ends with a larger amount of hydrogen than K04-B6 as a result 
of the major role played by photodisintegration reactions when the temperature
exceeds $\sim 2$ GK. This is illustrated in Fig. 2, where the time evolution
of the hydrogen mass-fraction is plotted for the two models K04-B6 (low T)
and K04-B7 (high T). Indeed, the {\it bump} exhibited by Model K04-B7 during the
20 s after $T_{peak}$ (that is, when the temperature ranges between
1.5 and 2.5 GK) is caused by the protons released through ($\gamma$, p) reactions
on a number of nuclear species.
Hence, the final hydrogen mass fraction may provide a diagnostic of the peak
temperature achieved during the explosion (for bursts of similar duration).

Concerning nuclear uncertainties, Model K04-B6 is characterized by 49 critical
reactions, the most important ones being 
$^{61}$Ga(p,$\gamma$), $^{82}$Zr(p,$\gamma$), $^{65}$As(p,$\gamma$), $^{86,87}$Mo(p,$\gamma$),  
$^{84}$Nb(p,$\gamma$), with a minor role played by $^{92}$Ru(p,$\gamma$)
(Table 9). For Model K04-B7, we find 53 critical reactions, with the largest roles played by $^{69}$Br(p,$\gamma$),
followed by $^{96}$Ag(p,$\gamma$), and $^{103}$In(p,$\gamma$) (Table 10).
 It is also worth mentioning that in both models only 2 $\alpha-induced$ reactions appear to
 be influential (one being $^{15}$O($\alpha$,$\gamma$)).

As expected, the most important $\beta$-decay rates involve heavier species in
Model K04-B7 than in K04-B6, as a result of the former's larger peak temperature (see Table 11). Hence, the $\beta-decay$  rates of $^{18}$Ne, $^{64}$Ge, $^{68}$Se, $^{72}$Kr, $^{76}$Sr, and $^{80}$Zr,
are influential in Model K04-B6, whereas
those of $^{68}$Se, $^{72}$Kr, $^{76}$Sr, $^{80}$Zr, $^{88}$Ru, $^{92}$Pd, $^{101, 102}$Sn,
are important in Model K04-B7 (again only for uncertainty factors of 10, up and down).

\section{Monte Carlo Simulations}

As discussed earlier, realistic simulations of Type I X-ray bursts have revealed a 
dramatic extension of the nuclear path, reaching the proton-drip line at many
points, and extending typically up to A=50--60 (Fisker et al. 2008), or eventually, 
up to the SnSbTe-mass region
(Schatz et al. 2001), or beyond (Koike et al. 2004).
The overall number of nuclear processes is huge, since 
many different reactions are involved in the synthesis and/or
destruction of any particular isotope during the explosion.

Recently, it has been claimed (Roberts et al. 2006) that, because of the coupling
of so many different channels in XRBs, {\it traditional} sensitivity studies (like the one
presented in Section 3), in which only one reaction is varied while the others
are held constant, cannot properly address all the important correlations between 
rate uncertainties and nucleosynthetic predictions, leading to wrong (or at least,
biased) conclusions.

Here, we will examine the impact of simultaneously varying all reaction
rates in the highly-coupled environment characteristic of an XRB through Monte Carlo 
simulations, and will discuss the feasibility of traditional sensitivity
studies through a direct comparison. 
   
\subsection{The Monte Carlo code}

The simultaneous variation of reaction rates relies on
the generation of a set of pseudo-random numbers that follow a given
probability density function.

Since nuclear reaction rates are positive quantities, we assume that
the random {\it enhancement factors} that are applied to each individual reaction
follow a log-normal probability function 
(note that other distributions, such as Gaussian, may give negative values).
The log-normal probability function adopted has the form:

\begin{equation}
f(x; \mu,\sigma)=\frac{1}{\sqrt{2\pi} \, \sigma \, x}\, 
exp [(-(ln(x) - \mu)^2/2 \sigma^2], \, x > 0
\end{equation} 

where $\mu$ and $\sigma$ are the mean and the standard
deviation of $ln(x)$, respectively\footnote{By definition, if $x$ is log-normally distributed, 
$ln(x)$  is normally distributed.  Note that
$\mu^* \equiv e^{\mu}$ is the geometric mean (as well as the median) 
of the log-normal distribution, whereas $\sigma^* \equiv e^{\sigma}$ is 
the geometric standard deviation.}.

 This log-normal distribution is implemented in such a way
that the geometric mean of all enhancement factors is $\mu^* $ = 1, and the probability to generate a 
random number 
between $0.1$ and $10$ (that is, in the range [$\mu^*/(\sigma^*)^2$, $\mu^* \times (\sigma^*)^2$]), 
to match the factors adopted for our sensitivity study in Section 3, is chosen to be 95.5\%.  
This is accomplished by setting $\sigma^* = \sqrt{10}$.  From the many algorithms
available in the literature for generating pseudo-random numbers, 
we have used that from Guardiola et al. (1995).

The set of enhancement factors is applied to all nuclear
reactions included in the network; the same enhancement factor was applied to a forward-reverse reaction 
pair\footnote{A rigorous Monte Carlo procedure would require a
random sampling of the reaction Q-value for the calculation of the reverse reaction rate.
We did not address this extra complication, which we leave to future investigations.}.
This is coupled to a 
post-processing nucleosynthesis code that allows us to compute
the final yields for a particular temperature and density versus time profile,
in a similar way as described in Sections 2 \& 3.
The process is then iterated for a pre-selected number of trials, in order
to achieve statistically sound results.
The series of computations performed in this work rely on 1000 trials.
 Although previous Monte Carlo simulations have been based on a larger
number of trials (10,000, in Smith et al. 2002 and Hix et al. 2003; 50,000, 
in Roberts et al. 2006), we feel our choice is sufficient for
the goals of this manuscript (see sections 4.2 and 5).
 
\subsection{Results}

We have applied the Monte Carlo technique to our set of models
discussed in Section 3. However, since results are qualitatively similar in all cases
 we will restrict our discussion here to Model K04 (Koike et al. 2004).

Before a thorough analysis of the results is made, several aspects must be considered in detail.  
First, the use of realistic uncertainties affecting all the rates is critical in Monte Carlo simulations, 
since the simultaneous variation of all reactions may otherwise result in inflated uncertainties 
in the final yields (Smith et al. 2002). Because of this, studies based on individual reaction-rate variations
are more easily interpreted.
This is illustrated in Figures 3 \& 4, where {\it relative abundances} and their {\it normalized
geometric standard deviations} (see Limpert et al. 2001, and Carobbi et al. 2003) 
are plotted as a function of the mass number, for Model K04. 
The {\it relative abundances} correspond to the geometric mean abundances obtained from the set of Monte Carlo trials 
normalized to abundances found with standard rates. In turn, the {\it normalized geometric standard deviations}  
define the interval for which the probability
to enclose any possible geometric mean abundance for a given species matches a certain value (95.5\%, in this study). 

The simultaneous variation of all nuclear processes in bulk between $0.1$ and $10$, according to our 
log-normal distribution,
results in large uncertainty bars for many isotopes (Figure 3). 
Notice, however, that these final abundance uncertainties are overestimated 
since all $\beta^+$-decays, as well as some
important, relatively well-known reactions, such as the triple-$\alpha$, were allowed to vary in this way.
As discussed in Section 3, more realistic uncertainties must be used instead, since 
these rates are known with better precision (usually better than $\sim$ 30\%, for $\beta$-decay rates). 
Indeed, when all nuclear processes, except the 3$\alpha$ and all the $\beta$-decay rates, are allowed to vary between 
roughly $0.1$ and $10$ in the Monte Carlo study, the overall uncertainties in the final yields decrease dramatically (Figure 4).
Similar results have been achieved for Model F08 (see Figures 5 \& 6).

A second warning associated with Monte Carlo simulations involves the reduced subset of reactions 
whose variation affects the overall energy production (see Section 5). 
In individual reaction-rate variation studies, these reactions can be appropiately flagged for  
separate, detailed analysis with better numerical tools (semi-analytical or hydrodynamic codes that can properly address 
changes in the temperature and density profiles driven by variations in the total energy output).
In Monte Carlo studies, however, one cannot simply remove those trials in which the
overall energy production is modified as this would affect the input distribution of enhancement 
factors (which are assumed to be random). 
Therefore, while results from individual reaction-rate variations are not corrupted by 
these effects, Monte Carlo simulations
cannot disentangle this from the overall analysis and hence the interpretation of the results 
has to be taken with caution.

Figure 4 shows the impact of the simultaneous variation of all rates (except for the 3$\alpha$ and all the 
$\beta$-decay rates) on the final yields, for the temperature and density versus 
time profiles of Model K04 (Koike et al. 2004). It illustrates the interplay of multiple nuclear processes in the
highly coupled environment of an XRB. The identification of the key reactions, whose uncertainties have the
largest impact on the final yields, is more complicated than in {\it traditional} sensitivity studies.
Figure 4 shows indeed which isotopes are mostly affected by the uncertainties associated with the rates.
However, the identification of those specific reactions that are perhaps most responsible for those changes 
is, by no means, straightforward. Following Smith et al. (2002), Hix et al. (2003), and Roberts et al. (2006), we have searched
for possible correlations between variations in the final abundance of a specific nucleus and each 
nuclear reaction rate that was varied in the Monte Carlo routine. 

In many cases, correlations are small, as illustrated in Figure 7: 
although the relative abundance of $^{69}$Ge shows a significant
variability, it cannot be attributed to the uncertainty affecting 
the $^{56}$Ni($\alpha$, p)$^{59}$Cu (and $^{59}$Cu(p,$\alpha$)$^{56}$Ni) rate, 
since the fit has a correlation coefficient of only r=0.101.
 In sharp contrast,
Figs. 8 \& 9 show a large correlation between the $^{69}$Ge abundance
and $^{69}$Se(p, $\gamma$)$^{70}$Br, as well as between $^{72}$Se and 
$^{65}$As(p, $\gamma$)$^{66}$Se. However,
despite the large correlation, varying the $^{65}$As(p, $\gamma$) rate
by a factor of 10 has no dramatic impact on the final $^{72}$Se yield (Fig. 9).
Hence, a key reaction in the Monte Carlo approach must have a significant 
correlation with the yield of an isotope (as indicated by the correlation coefficient)
as well as a pronounced impact on the final abundance of an isotope as its rate is varied.

The most important correlations
from the Monte Carlo study applied to Model K04, as 
characterized by fits with correlation coefficients $r > 0.5$, are listed in Table 15. 
From the 50 reactions listed in the table, all but 5 (namely, $^{34}$Ar(p, $\gamma$), $^{53}$Co(p, $\gamma$), 
$^{62}$Ga(p, $\gamma$), $^{69}$Br(p, $\gamma$),
and  $^{80}$Y(p, $\gamma$)) were previously identified in the study
based on individual reaction-rate variations (Table 1). 

Table 16 displays reactions
that have the largest effects on the final yields for Model K04, according to
the individual reaction-rate variation study (column 2) and Monte Carlo simulations (column 3).
Reactions included in Table 16 were restricted to those that, in either the individual
reaction-rate variation study or in the Monte Carlo simulations, affected the final
abundances of the isotopes by at least a factor of 2 (for the former, see Table 1; for the
latter, reactions from Table 15 without brackets in the slope column).
Qualitatively, results are in excellent agreement, except 
at the very end of the network (nuclei with A $>$ 97).  
Indeed, most of the discrepancies correspond to reactions that were indeed identified in Table 15, but
that affect the final abundances by less than a factor of 2 when enhancement factors 
ranging between 0.1 and 10 are considered.
These discrepancies might be related to the correlations
that the individual reaction-rate variation study are not capable to deal with.

Similar agreement is reported on the comparison of both techniques when applied to Model F08
(Fisker et al. 2008), as illustrated in Tables 17 \& 18.

Finally, we would like to stress that our preliminary Monte Carlo studies rely on 1,000 trials, a limited number
that calls for a justification. To this end, we have extended the study of Model K04 to 4,200
trials. Indeed, no significant differences, neither
in the list of key reactions, nor on the correlation coefficients and slopes have been found.

\section{Discussion and Conclusions}

Table 19 summarizes the most important reactions collected from Tables 1-10. 
For the sake of brevity we have restricted Table 19 and the dicussion here to those reactions that affect the yield 
of 3 species or more in any of our 10 models, in the individual-variation studies.
We have also carefully identified reactions that were seen to 
modify the XRB energy output when their rates were varied by a factor of 10, up or down.
 This is explicitly indicated in Table 20, which lists
the subset of reactions with any impact on XRB yields (reactions identified in Tables 1 to 10, that affect, 
at least, one isotope) that simultaneously modify the overall energy output by more than 5\% at some point during
the burst, when their nominal rates are varied by a factor of 10, up or down. This table has to be taken
as a warning of the limitations of post-processing techniques (both for individual-variation and Monte Carlo studies).
Furthermore, we have identified some additional reactions in our studies which affect the energy output,
but remarkably did not affect any yields in any of our models (for instance, $^{14}$O($\alpha$, p), 
$^{27}$Si(p, $\gamma$), $^{31}$S(p, $\gamma$), or $^{35}$S(p, $\gamma$)).
Several aspects are worth noting here. First, the total number of reactions affecting the energy output, for any model, is small.
Second, as indicated in Table 20, some of those reactions are known with better precision than a factor of 10.
Third, there is no way to overcome this problem in the context of post-processing calculations. Indeed,
a self-consistent analysis with a hydrodynamic code capable of self-adjusting both the temperature and the density
of the stellar envelope seems mandatory to address this issue, for the few reactions of concern. And fourth, the presence of 
reactions affecting the energy output is particularly dangerous within a Monte Carlo context, as results
rely on the simultaneous variation of all rates, whereas most
of the results obtained from an individual reaction-rate variation approach remain unaffected.
Note, however that Tables 19 \& 20 reveal a very limited number of reactions that were seen to affect the energy output for 
Models K04 and F08, giving support to the results obtained in our Monte-Carlo studies.  

It is also worth mentioning that no discussion involving the triple-$\alpha$ reaction or any beta-decay rate has been
made here, since these reactions, when varied within
realistic uncertainty limits, have no effect, neither on yields, nor on the nuclear energy output. 
The same applies to Q-value variations, hence results listed in Tables 13 \& 14 are not affected by 
variations of the energy output.

\subsection{Overview of the most influential reactions}

The main purpose of this Section is to 
elaborate on the reactions listed in Tables 19 \& 20, namely, the most influential reactions found in this study. 

First, we will focus on those rates that were drawn from theoretical estimates, due to insufficient or unavailable 
experimental information. Their main characteristics are summarized in Table 21.  
To help illustrate systematic uncertainities associated with these important reactions, we have compared the 
theoretical rates adopted in our network with those found in the recent REACLIBv0 compilation\footnote{http://www.nscl.msu.edu/~nero/db/}
over the temperature range covered in our studies.  Overall, the agreement is quite good; however, we note that even when 
using the same basic Hauser-Feshbach code (NON-SMOKER) along with similar Q-values, a difference in the rates as large 
as a factor of $\sim$3 is obtained.  Differences as large as a factor of $\sim$4 are seen between rates from different 
statistical model codes, but using similar Q-values.  Moreover, the factor of 10 disagreement for the $^{61}$Ga(p, $\gamma$) 
rate arises from the comparison between a NON-SMOKER result and a shell-model calculation.  
The magnitude of these discrepancies lends support to our choice of varying rates by a factor of 10 rather 
than by a significantly larger factor.

We will now focus on the rates listed in Tables 19 \& 20 that have been determined experimentally, and we will assess 
whether a factor of 10 variation is reasonable for these rates.  For those cases where a smaller uncertainty is justified, 
we have performed additional post-processing calculations to supplement results from Tables 1-10 and 20 (namely, to determine 
the impact of individual rate variations by smaller factors on yields and on the overall nuclear energy outuput).  For reference, we 
will continue to compare the rate adopted in this work with that used in the REACLIBv0 compilation.
Although we do not discuss experimental information (if available) for other reactions listed in Tables 1-10 (namely, those that 
affected less than 3 isotopes in any model), this must of course be considered by anyone examining reactions beyond the most 
influential ones listed in Table 19.  

\begin{itemize}
\item $^{12}$C($\alpha$, $\gamma$)$^{16}$O:
We have used the Kunz et al. (2002) experimental rate which, over the range of 
temperatures used in our models, agrees to better than 20\% with the recommended NACRE 
(Angulo et al. 1999) rate, and to better than a factor of $\sim$2 with the recommended 
Buchmann (1996) rate adopted in REACLIBv0.  A factor of $\sim$3 variation in 
our rate would cover the limits given in both NACRE and Buchmann (1996).  Hence, we have 
varied this rate individually by a factor of 3, up and down, and tested its effect for 
those models in which a factor of 10 variation, as discussed in Section 3, had an 
impact (Models F08, K04-B2, K04-B4, and K04-B5).  Indeed, only Model K04-B2 
(where the $^{12}$C yield is affected by a factor 0.47 when this rate is multiplied by 3) 
and K04-B5 (where $^{20}$Ne and $^{24}$Mg are affected by factors 0.44 and 0.47, respectively, 
when the rate is reduced by a factor 3) reveal changes in the final yields (note that 
variation of this reaction by a factor of 10 did not affect the overall nuclear energy output in any of our models). 

\item $^{18}$Ne($\alpha$, p)$^{21}$Na:
We use the Chen et al. (2001) experimental rate, which agrees to a factor $\sim$3 with the G\"orres et al. (1995) 
Hauser-Feshbach (SMOKER) calculation adopted in REACLIBv0.  There is, however, additional data from Groombridge et al. (2002).  
Indeed, using the information from Chen et al. (2001), for $E_r < 1.7$ MeV, and that from Groombridge et al. (2002), 
for $E_r > 1.98$ MeV, we find a rate which deviates from the Chen et al. rate by 30\%, a factor of $\sim$3 and a factor 
of $\sim$7 at 1.0, 1.4, and 2 GK, respectively.  Since we found this reaction to affect yields and/or the total nuclear 
energy in Models K04-B1 and K04-B6, neither of which reach temperatures above 1.36 GK, we have restricted
our analysis to a variation of the Chen et al. rate by a factor of 3, up and down. 
No impact  on the yields is found when varying the rate as such; however, Model K04-B1 shows some variation of
the total energy output when the rate is multiplied by 3.

\item  $^{15}$O($\alpha$, $\gamma$)$^{19}$Ne: 
We use the Davids et al. (2003) rate, which agrees to factor of $\sim$3 with the Hahn et al. (1996) rate used in REACLIBv0, 
over the temperatures spanned by our models.  
If we combine the new information reported by Tan et al. (2007), for $E_x < 4.55$ MeV, with the information in Davids et al. (2003),
for the states at $E_x$ = 4.600, 4.712 and 5.092 MeV, we calculate a new rate that agrees to a factor $\sim$2 with the original 
Davids et al. rate, over the relevant temperatures. 
Varying our rate within a factor $\sim$3 would cover the uncertainty limits in the Tan-Davids rate calculation.  
We have tested the impact of varying our $^{15}$O($\alpha$, $\gamma$) rate by a factor 3 up and down,
for Models K04, K04-B1, and K04-B6.  We find that 
the total nuclear energy is affected only in the early stages of Model K04, when the rate is multiplied by 3. Concerning
the yields, only $^{15}$N is affected (consistent with Tables 1, 4 and 9): when the rate is reduced by a factor 3 the 
$^{15}$N normalized yields in Models K04, K04-B1 and K04-B6 are 3.13, 2.75, and 3.17, respectively; when the rate is increased by a 
factor 3, no effect is seen except in Model K04-B1, where the normalized yield of $^{15}$N decreases to 0.35.      

\item $^{23}$Al(p, $\gamma$)$^{24}$Si:
We use the Schatz et al. (1997) rate, as does REACLIBv0.  This rate is based on both theoretical estimates and measurements 
of excited states in $^{24}$Si.  Its uncertainty spans up to 3 orders of magnitude, for typical XRB temperatures. For this 
reason, we deem our results from varying this rate by a factor 10 to be adequate.    

\item $^{24}$Mg($\alpha$, p)$^{27}$Al:
Both REACLIBv0 and our network rely on the experimental rate reported in Iliadis et al. (2001).  Above 0.3 GK, this rate is reported 
to be uncertain by only $\pm$20\%.  No effect on the yields is found when the rate is varied by 20\% in our post-processing 
calculations (Model K04-B2; see Table 20). The total nuclear energy is, however, affected when this rate is increased by 20\%, in 
that particular model.

\item $^{26g}$Al(p, $\gamma$)$^{27}$Si:
We use the (unpublished) rate from the PhD thesis of Vogelaar (1989), as does REACLIBv0 (the new TRIUMF measurement 
--Ruiz et al. (2006)-- on the 184 keV resonance does not affect the rate at typical XRB temperatures).  According to NACRE, 
the experimental information is only sufficient to determine the rate for $T < 0.9$ GK (they use Hauser-Feshbach calculations 
to extend this rate to higher temperatures).  Our rate agrees with the NACRE recommended rate to 30\% over the temperature 
range covered by our models; varying our rate by a factor of 2 up and down would cover the uncertainties reported in NACRE.
Assuming such a degree of uncertainty in our rate for Model F08 (see Table 20) leads to no effect on any yield, 
nor on the total nuclear energy.

\item $^{28}$Si($\alpha$, p)$^{31}$P:
The rate used in our network and that of REACLIBv0 is from the compilation of Iliadis et al. (2001).  An uncertainty of $\pm 20\%$ is 
assigned to this rate above 0.2 GK.  Actually, varying this rate by $\pm$ 20\% in Model K04-B4 (see Table 20) has no effect on 
any yields, nor on the total nuclear energy.

\item $^{32}$S($\alpha$, p)$^{35}$Cl:
We use the Iliadis et al. (2001) experimental rate, as does REACLIBv0.  The uncertainty of this rate, over XRB typical temperatures,
spans up to 3 orders of magnitude (Iliadis et al. 2001). Consequently, we feel that our results, based on a variation  of this rate
by a factor of 10 (see Table 20), are reliable.

\item $^{35}$Cl(p, $\gamma$)$^{36}$Ar:
Again, we adopt the Iliadis et al. (2001) experimental rate, as does REACLIBv0.  The uncertainty in this rate is $\sim$20\% above 0.2 GK
(Iliadis et al. 2001). We find no effect on any yield when varying this rate by $\pm$20\% in our post-processing calculations 
for Model K04-B2 (see Table 20). However, we remarkably find that increasing this rate by 20\% does indeed affect the total 
nuclear energy at the early stages of the TNR in this model.

\end{itemize}

Finally, because of past interest in the literature, we discuss some specific reactions that we found to affect the 
total nuclear energy when their rates were varied individually by a factor of 10, but affected no yields whatsoever 
(see Section 5.1): $^{31}$S(p, $\gamma$), $^{35}$Ar(p, $\gamma$), $^{27}$Si(p, $\gamma$), and $^{14}$O($\alpha$, p).  

\begin{itemize}
\item $^{31}$S(p, $\gamma$)$^{32}$Cl, $^{35}$Ar(p, $\gamma$)$^{36}$K, and $^{27}$Si(p, $\gamma$)$^{28}$P:
For these 3 reactions, we have used the rates reported in Iliadis et al. (1999),  determined through the use of experimental 
information when available (e.g., excitation energies), along with calculations and information from the respective mirror nuclei.  
Judging from the uncertainties presented in that work, we conclude that uncertainty factors of 2, 3 and 2 (up and down) are 
more reasonable for the $^{31}$S(p, $\gamma$), $^{35}$Ar(p, $\gamma$), and $^{27}$Si(p, $\gamma$) rates over XRB temperatures, respectively.  
Accordingly, we individually varied each of these reactions within those limits for the models affected (i.e., 
Models K04-B1 and K04-B7 for $^{31}$S(p, $\gamma$); Models K04-B2 and K04-B7 for $^{35}$Ar(p, $\gamma$); 
and Models K04-B4 and K04-B7 for $^{27}$Si(p, $\gamma$)) to determine the resulting impact on the total nuclear energy,
as no yields were affected by varying any of these by a factor of 10. Increasing the $^{31}$S(p, $\gamma$) rate by a factor of 2 
changed the total nuclear energy (by at least 5\% at some point of the burst) in Model K04-B1; 
increasing the $^{35}$Ar(p, $\gamma$) rate by a factor 3 changed the nuclear energy in Model K04-B2; and increasing the 
$^{27}$Si(p, $\gamma$) rate by a factor 2 changed the nuclear energy in Model K04-B7.  The impact of these reactions on XRB 
light curves has been examined in Iliadis et al. (1999) and in Thielemann et al. (2001).  

\item $^{14}$O($\alpha$, p)$^{17}$F:
This reaction is of critical importance for breakout from the hot CNO-cycle.  We use the rate from Blackmon et al. (2003), 
which is larger than the Hahn et al. (1996) rate, adopted by REACLIBv0, by a factor of $\sim$10 at typical XRB temperatures. 
This increased rate is due to the inclusion of some of the contributions from the $^{17}$F* exit channel by Blackmon et al.;
branches for states of $E_x > 7$ MeV in $^{18}$Ne were not measured though.  Actually, Notani et al. (2004a, b) observed what could 
be the population of $^{17}$F* through a state at $E_x$($^{18}$Ne) $\sim 7.1$ MeV, but the interpretation of their 
results has been questioned by Fu et al. (2007).  As the studies of Blackmon et al. and Notani et al. are both published in 
only preliminary forms, and taking into account the argument of Fu et al., we  find it difficult to evaluate the uncertainty in 
this rate.  Variations by a factor of 10 affected nuclear energy in 5 of our 10 models: K04, F08, K04-B1, K04-B3 and K04-B4.  
Further efforts to constrain this rate, based on analyses of previous measurements and/or new measurements, would certainly be desirable.    
   
\end{itemize}

In summary, we have identified a very limited number of reactions (see Tables 19--21 and this section) that play a significant 
role in our XRB nucleosynthesis studies.  Indeed, our results can help to guide and motivate future measurements by 
experimental nuclear physicists.  Stellar modelers, as well, may tackle the challenge to properly address the role played by 
the few reactions flagged as affecting the overall energy output, an aspect that lies beyond the possibilities 
offered by post-processing calculations.

Finally, we have compared the two general approaches to the sensitivity study: individual reaction-rate variations, and 
simultaneous variation of all rates through Monte Carlo techniques.  The highly coupled environment characteristic of an XRB 
provides the opportunity to test for the effects of correlations in the uncertainties of reaction rates.  We find similar 
results from both approaches, with minor differences attributed to such correlation effects.  

\acknowledgments

This work has been partially supported by the Spanish MEC grant
AYA2007-66256, by the E.U. FEDER funds, and by the U.S. Department of Energy under Contract No. DE-FG02-97ER41041.

\clearpage


\begin{deluxetable}{lccc}
\small
\tablecaption{Final abundance ratios, $X_i/X_{i,std}$, for Model K04 (Koike et al. 2004), resulting from
                  reaction-rate variations by a factor of 10 (up and down).}
\tablehead{ 
\colhead{Reaction}   & 
\colhead{Isotope}   & 
\colhead{10} & 
\colhead{0.1} 
}
\startdata
$^{15}$O($\alpha$, $\gamma$)$^{19}$Ne& $^{15}$N  &  ...  &   8.30 \\
$^{18}$Ne($\alpha$, p)$^{21}$Na    & $^{15}$N    &  0.43 &   ...  \\
                                   & $^{18}$F    &  0.43 &   ...  \\
$^{31}$Cl(p, $\gamma$)$^{32}$Ar    & $^{30}$Si   &  0.44 &   ...  \\
$^{43}$V(p, $\gamma$)$^{44}$Cr     & $^{44}$Ti   &  ...  &   0.49 \\
$^{47}$Mn(p, $\gamma$)$^{48}$Fe    & $^{46}$Ti   &  0.31 &   ...  \\
$^{56}$Ni(p, $\gamma$)$^{57}$Cu    & $^{56}$Ni   &  ...  &   7.78 \\
$^{56}$Cu(p, $\gamma$)$^{57}$Zn    & $^{55}$Co   &  ...  &   2.64 \\
\enddata
\end{deluxetable}

\begin{deluxetable}{lccc}
\small
\tablecaption{Same as Table 1, for Model S01 (Schatz el at. 2001).}
\tablehead{ 
\colhead{Reaction}   & 
\colhead{Isotope}   & 
\colhead{10} & 
\colhead{0.1} 
}
\startdata
$^{15}$O($\alpha$, $\gamma$)$^{19}$Ne& $^{15}$N  &  ...  &   8.84 \\
$^{52}$Fe(p, $\gamma$)$^{53}$Co    & $^{52}$Fe   &  ...  &   8.48 \\
$^{52}$Co(p, $\gamma$)$^{53}$Ni    & $^{52}$Fe   &  ...  &   2.03 \\
$^{56}$Ni(p, $\gamma$)$^{57}$Cu    & $^{56}$Ni   &  ...  &   8.33 \\
$^{56}$Ni($\alpha$, p)$^{59}$Cu    & $^{1}$H     &  0.36 &   ...  \\
                                   & $^{52}$Fe   &  2.40 &   ...  \\
                                   & $^{56}$Ni   &  2.31 &   ...  \\
                                   & $^{69}$Ge   &  2.27 &   ...  \\
\enddata
\end{deluxetable}

\begin{deluxetable}{lccc}
\small
\tablecaption{Same as Table 1, for Model F08 (Fisker el at. 2008).}
\tablehead{ 
\colhead{Reaction}   & 
\colhead{Isotope}   & 
\colhead{10} & 
\colhead{0.1} 
}
\startdata
$^{12}$C(p, $\gamma$)$^{13}$N      &  $^{12}$C   &  0.41 &   ...  \\
$^{12}$C($\alpha$, $\gamma$)$^{16}$O& $^{12}$C   &  0.42 &   ...  \\
                                   &  $^{16}$O   &  3.27 &   0.36 \\
                                   &  $^{20}$Ne  &  3.03 &   0.42 \\
                                   &  $^{24}$Mg  &  2.40 &   ...  \\
$^{16}$O($\alpha$, $\gamma$)$^{20}$Ne& $^{16}$O  &  0.11 &   8.36 \\
$^{20}$Ne($\alpha$, $\gamma$)$^{24}$Mg& $^{20}$Ne&  0.10 &   9.84 \\
$^{22}$Mg($\alpha$, p)$^{25}$Al    &  $^{25}$Mg  &  ...  &   8.81 \\
\enddata
\end{deluxetable}

\begin{deluxetable}{lccc}
\small
\tablecaption{Same as Table 1, for Model K04-B1.}
\tablehead{ 
\colhead{Reaction}   & 
\colhead{Isotope}   & 
\colhead{10} & 
\colhead{0.1} 
}
\startdata
$^{15}$O($\alpha$, $\gamma$)$^{19}$Ne& $^{15}$N  &  0.11 &   6.14 \\
$^{18}$Ne($\alpha$, p)$^{21}$Na    &  $^{15}$N   &  0.28 &   ...  \\
                                   &  $^{18}$F   &  0.29 &   ...  \\
                                   &  $^{21}$Ne  &  0.26 &   2.03 \\
$^{22}$Mg(p, $\gamma$)$^{23}$Al    &  $^{22}$Na  &  0.49 &   6.21 \\
$^{23}$Al(p, $\gamma$)$^{24}$Si    &  $^{22}$Na  &  ...  &   4.45 \\
$^{27}$P(p, $\gamma$)$^{28}$S      &  $^{27}$Al  &  0.25 &   6.34 \\
$^{31}$Cl(p, $\gamma$)$^{32}$Ar    &  $^{30}$Si  &  0.16 &   2.57 \\
\enddata
\end{deluxetable}

\begin{deluxetable}{lccc}
\small
\tablecaption{Same as Table 1, for Model K04-B2.}
\tablehead{ 
\colhead{Reaction}   & 
\colhead{Isotope}   & 
\colhead{10} & 
\colhead{0.1} 
}
\startdata
$^{12}$C($\alpha$, $\gamma$)$^{16}$O& $^{12}$C   &  0.19 &   2.37 \\
                                   &  $^{24}$Mg  &  ...  &   0.26 \\
                                   &  $^{28}$Si  &  ...  &   0.29 \\
$^{20}$Ne($\alpha$, $\gamma$)$^{24}$Mg& $^{20}$Ne&  ...  &   8.28 \\
$^{24}$Mg($\alpha$, $\gamma$)$^{28}$Si& $^{24}$Mg&  0.16 &   3.77 \\
                                   &    $^{34}$S &  0.48 &   4.30 \\
$^{24}$Mg($\alpha$, p)$^{27}$Al    &    $^{34}$S &  3.99 &   0.49 \\
$^{28}$Si($\alpha$, $\gamma$)$^{32}$S & $^{28}$Si&  0.31 &   2.63 \\
\enddata
\end{deluxetable}

\begin{deluxetable}{lccc}
\small
\tablecaption{Same as Table 1, for Model K04-B3.}
\tablehead{ 
\colhead{Reaction}   & 
\colhead{Isotope}   & 
\colhead{10} & 
\colhead{0.1} 
}
\startdata
$^{15}$O($\alpha$, $\gamma$)$^{19}$Ne& $^{15}$N  &  ...  &   8.17 \\
$^{18}$Ne($\alpha$, p)$^{21}$Na    &  $^{15}$N   &  0.43 &   ...  \\
                                   &  $^{18}$F   &  0.44 &   ...  \\
$^{31}$Cl(p, $\gamma$)$^{32}$Ar    &  $^{30}$Si  &  0.45 &   ...  \\
$^{43}$V(p, $\gamma$)$^{44}$Cr     &  $^{42}$Ca  &  ...  &   4.68 \\
                                   &  $^{44}$Ti  &  ...  &   0.48 \\
$^{47}$Mn(p, $\gamma$)$^{48}$Fe    &  $^{46}$Ti  &  0.32 &   ...  \\
$^{56}$Ni(p, $\gamma$)$^{57}$Cu    &  $^{56}$Ni  &  ...  &   7.93 \\
\enddata
\end{deluxetable}

\begin{deluxetable}{lccc}
\small
\tablecaption{Same as Table 1, for Model K04-B4.}
\tablehead{ 
\colhead{Reaction}   & 
\colhead{Isotope}   & 
\colhead{10} & 
\colhead{0.1} 
}
\startdata
$^{12}$C($\alpha$, $\gamma$)$^{16}$O& $^{12}$C   &  0.24 &   2.53 \\
                                   &  $^{16}$O   &  2.38 &   ...  \\
                                   &  $^{20}$Ne  &  2.30 &   0.26 \\
                                   &  $^{24}$Mg  &  ...  &   0.29 \\
                                   &  $^{28}$Si  &  ...  &   0.41 \\
$^{16}$O($\alpha$, $\gamma$)$^{20}$Ne& $^{16}$O  &  ...  &   7.65 \\
$^{20}$Ne($\alpha$, $\gamma$)$^{24}$Mg& $^{20}$Ne&  ...  &   7.87 \\
$^{24}$Mg($\alpha$, $\gamma$)$^{28}$Si& $^{24}$Mg&  0.20 &   3.48 \\
\enddata
\end{deluxetable}

\begin{deluxetable}{lccc}
\small
\tablecaption{Same as Table 1, for Model K04-B5.}
\tablehead{ 
\colhead{Reaction}   & 
\colhead{Isotope}   & 
\colhead{10} & 
\colhead{0.1} 
}
\startdata
$^{12}$C($\alpha$, $\gamma$)$^{16}$O& $^{12}$C   &  0.26 &   ...  \\
                                   &  $^{16}$O   &  2.54 &   ...  \\
                                   &  $^{20}$Ne  &  2.44 &   0.18 \\
                                   &  $^{24}$Mg  &  2.13 &   0.21 \\
                                   &  $^{28}$Si  &  ...  &   0.46 \\
$^{16}$O($\alpha$, $\gamma$)$^{20}$Ne& $^{16}$O  &  ...  &   7.66 \\
$^{20}$Ne($\alpha$, $\gamma$)$^{24}$Mg& $^{20}$Ne&  ...  &   7.87 \\
$^{24}$Mg($\alpha$, $\gamma$)$^{28}$Si& $^{24}$Mg&  0.20 &   3.38 \\
\enddata
\end{deluxetable}

\begin{deluxetable}{lccc}
\small
\tablecaption{Same as Table 1, for Model K04-B6.}
\tablehead{ 
\colhead{Reaction}   & 
\colhead{Isotope}   & 
\colhead{10} & 
\colhead{0.1} 
}
\startdata
$^{15}$O($\alpha$, $\gamma$)$^{19}$Ne& $^{15}$N  &  ...  &   9.47 \\
$^{18}$Ne($\alpha$, p)$^{21}$Na    &  $^{15}$N   &  0.42 &   ...  \\
                                   &  $^{18}$F   &  0.41 &   ...  \\
$^{22}$Mg(p, $\gamma$)$^{23}$Al    &  $^{22}$Na  &  ...  &   3.14 \\
$^{23}$Al(p, $\gamma$)$^{24}$Si    &  $^{22}$Na  &  ...  &   5.61 \\
$^{27}$P(p, $\gamma$)$^{28}$S      &  $^{27}$Al  &  ...  &   4.73 \\
                                   &  $^{28}$Si  &  ...  &   0.48 \\
$^{39}$Ca(p, $\gamma$)$^{40}$Sc    &  $^{39}$K   &  ...  &   8.52 \\
\enddata
\end{deluxetable}

\begin{deluxetable}{lccc}
\small
\tablecaption{Same as Table 1, for Model K04-B7.}
\tablehead{ 
\colhead{Reaction}   & 
\colhead{Isotope}   & 
\colhead{10} & 
\colhead{0.1} 
}
\startdata
$^{15}$O($\alpha$, $\gamma$)$^{19}$Ne& $^{15}$N  &  ...  &   7.89 \\
$^{18}$Ne($\alpha$, p)$^{21}$Na    &  $^{18}$F   &  0.48 &   ...  \\
$^{31}$Cl(p, $\gamma$)$^{32}$Ar    &  $^{30}$Si  &  ...  &   2.17 \\
$^{34}$Ar(p, $\gamma$)$^{35}$K     &  $^{34}$S   &  0.37 &   2.14 \\
$^{47}$Mn(p, $\gamma$)$^{48}$Fe    &  $^{46}$Ti  &  ...  &   3.23 \\
$^{57}$Cu(p, $\gamma$)$^{58}$Zn    &  $^{57}$Ni  &  ...  &   2.12 \\
$^{61}$Ga(p, $\gamma$)$^{62}$Ge    &  $^{60}$Ni  &  0.13 &   5.21 \\
$^{65}$As(p, $\gamma$)$^{66}$Se    &  $^{64}$Zn  &  0.38 &   ...  \\
\enddata
\end{deluxetable}

\begin{deluxetable}{l}
\small
\tablecaption{Important $\beta$-decay (and $\beta$-delayed proton decay) rates, 
              as found in our ten models. See text for details.}
\tablehead{ 
\colhead{Reaction}   
}
\startdata
\hline
Model K04 \\
\hline
$^{64}$Ge($\beta^+$)$^{64}$Ga \\
$^{68}$Se($\beta^+$)$^{68}$As \\
$^{72}$Kr($\beta^+$)$^{72}$Br \\
$^{76}$Sr($\beta^+$)$^{76}$Rb \\
$^{80}$Zr($\beta^+$)$^{80}$Y  \\
$^{88}$Ru($\beta^+$)$^{88}$Tc \\
$^{92}$Pd($\beta^+$)$^{92}$Rh \\
\enddata
\end{deluxetable}

\begin{deluxetable}{lcc}
\small
\tablecaption{List of reactions selected for the Q-value variation study, restricted to
 proton-capture reactions on $A < 80$ nuclei, with $\left|Q\right|<$ 1 MeV, 
 and a Q-value uncertainty $>$ 50 keV 
 (see Audi et al. 2003a, and Section 3). }
\tablehead{ 
\colhead{Reaction}   & 
\colhead{Q-value\tablenotemark{a} (keV) }   & 
\colhead{Uncertainty (keV)}  
}
\startdata
$^{21}$Mg(p, $\gamma$)$^{22}$Al   &   17 &  95\\
$^{24}$Si(p, $\gamma$)$^{25}$P    & -828 & 197\\
$^{25}$Si(p, $\gamma$)$^{26}$P    &  140 & 196\\
 $^{25}$P(p, $\gamma$)$^{26}$S    &  191 & 357\\
 $^{26}$P(p, $\gamma$)$^{27}$S    &  719 & 281\\
 $^{29}$S(p, $\gamma$)$^{30}$Cl   & -314 & 202\\
 $^{30}$S(p, $\gamma$)$^{31}$Cl\tablenotemark{b}   &  294 &  50\\
$^{30}$Cl(p, $\gamma$)$^{31}$Ar   &  439 & 284\\
\enddata
\end{deluxetable}

\begin{deluxetable}{lccc}
\small
\tablecaption{Final abundance ratios, $X_i/X_{i,std}$, for Model K04 (Koike et al. 2004), resulting from
                  variations on Q-values (see Table 12, and text).}
\tablehead{ 
\colhead{Reaction}   & 
\colhead{Isotope}   & 
\colhead{High-Q (Q+$\Delta$Q)} & 
\colhead{Low-Q (Q-$\Delta$Q)} 
}
\startdata
$^{26}$P(p, $\gamma$)$^{27}$S      &  $^{25}$Mg  &  0.35 &   ...  \\
$^{46}$Cr(p, $\gamma$)$^{47}$Mn    &  $^{46}$Ti  &  0.23 &   ...  \\
$^{55}$Ni(p, $\gamma$)$^{56}$Cu    &  $^{55}$Co  &  ...  &   3.95 \\
$^{60}$Zn(p, $\gamma$)$^{61}$Ga    &  $^{60}$Ni  &  0.47 &   ...  \\
$^{64}$Ge(p, $\gamma$)$^{65}$As    &  $^{64}$Zn  &  0.08 &   4.13 \\
                                   &  $^{65}$Zn  &  .... &   3.04 \\
                                   &  $^{66}$Ge  &  .... &   3.05 \\ 
                                   &  $^{67}$Ga  &  .... &   3.65 \\
\enddata
\end{deluxetable}

\begin{deluxetable}{lccc}
\small
\tablecaption{Same as Table 13, for Model F08.}
\tablehead{ 
\colhead{Reaction}   & 
\colhead{Isotope}   & 
\colhead{High-Q (Q+$\Delta$Q)} & 
\colhead{Low-Q (Q-$\Delta$Q)} 
}
\startdata
$^{45}$Cr(p, $\gamma$)$^{46}$Mn    &  $^{45}$Ti  &  0.50 &   ...  \\
\enddata
\end{deluxetable}

\begin{deluxetable}{lccc}
\small
\tablecaption{Correlations between final yields and enhancement factors applied to reaction rates
        in the Monte Carlo approach, for Model K04. All correlations with $ r > 0.5$ (see text)
   are included. Reactions that have an impact of less than a factor of 2 on the final yields
   of correlated isotopes (for enhancement factors ranging between roughly 0.1 and 10) are indicated
   by brackets in the last column, in which slopes of the different linear fits 
   (final isotopic yield vs. rate enhancement factor, see e.g. figs. 8 and 9) are given. The error associated
   with the slopes is typically $\sim$1 -- 10\%.}
\tablehead{ 
\colhead{Isotope}   & 
\colhead{Reaction}   & 
\colhead{Correlation coefficient, r}  &
\colhead{Slope}
}
\startdata
$^{15}$N    & $^{15}$O($\alpha$, $\gamma$)$^{19}$Ne &  -0.778 &  -0.927 \\
$^{30}$Si   &     $^{31}$Cl(p, $\gamma$)$^{32}$Ar   & -0.751  &  [-0.205] \\
$^{33}$S    &     $^{18}$Ne($\alpha$, p)$^{21}$Na  & -0.601   &  [-0.063] \\
$^{34}$S    &     $^{34}$Ar(p, $\gamma$)$^{35}$K    & -0.603  &  [-0.148]   \\ 
$^{37}$Ar   &     $^{18}$Ne($\alpha$, p)$^{21}$Na  & -0.649   &  [-0.061]   \\
$^{38}$Ar   &     $^{18}$Ne($\alpha$, p)$^{21}$Na  & -0.571   &  [-0.062]    \\
$^{44}$Ti   &     $^{43}$V(p, $\gamma$)$^{44}$Cr    &  0.648  &  [0.134]    \\
$^{46}$Ti   &     $^{47}$Mn(p, $\gamma$)$^{48}$Fe   & -0.822  &  -0.292     \\
\enddata
\end{deluxetable}
\clearpage

\begin{deluxetable}{lcc}
\small
\tablecaption{Isotopes most affected by nuclear uncertainties, 
according to our sensitivity studies.
    Results from the individual reaction-rate approach (Table 1) are compared with
  those from Monte Carlo simulations (Table 15) for Model K04 (see text for details).} 
\tablehead{ 
\colhead{Isotope}   & 
\colhead{Reaction (Individual variations)}   & 
\colhead{Reaction (Monte Carlo)}    
}
\startdata
 $^{15}$N & $^{15}$O($\alpha$, $\gamma$)$^{19}$Ne &  $^{15}$O($\alpha$, $\gamma$)$^{19}$Ne    \\ 
          & $^{18}$Ne($\alpha$, p)$^{21}$Na       &  --  \\
 $^{18}$F &  $^{18}$Ne($\alpha$, p)$^{21}$Na      &  --  \\  
 $^{30}$Si & $^{31}$Cl(p, $\gamma$)$^{32}$Ar      &  --      \\ 
 $^{33}$S  &   --                                 &  --      \\
 $^{44}$Ti & $^{43}$V(p, $\gamma$)$^{44}$Cr       &  --       \\
 $^{46}$Ti & $^{47}$Mn(p, $\gamma$)$^{48}$Fe      &  $^{47}$Mn(p, $\gamma$)$^{48}$Fe        \\
 $^{55}$Co & $^{56}$Cu(p, $\gamma$)$^{57}$Zn      &   $^{56}$Cu(p, $\gamma$)$^{57}$Zn       \\
\enddata
\end{deluxetable}

\begin{deluxetable}{lccc}
\small
\tablecaption{Same as Table 15, for Model F08. }
\tablehead{ 
\colhead{Isotope}   & 
\colhead{Reaction}   & 
\colhead{Correlation coefficient, r}  &
\colhead{Slope}
}
\startdata
$^4$He    &   $^{12}$C($\alpha$, $\gamma$)$^{16}$O     &     -0.679  & [-0.030] \\
$^{12}$C  &   $^{12}$C($\alpha$, $\gamma$)$^{16}$O     &     -0.585  & [-0.203] \\
          &   $^{12}$C(p, $\gamma$)$^{13}$N            &     -0.581  & [-0.203] \\
$^{16}$O  &   $^{16}$O($\alpha$, $\gamma$)$^{20}$Ne    &     -0.878  & -0.945 \\
$^{20}$Ne &   $^{20}$Ne($\alpha$, $\gamma$)$^{24}$Mg   &     -0.894  & -1.01 \\ 
$^{24}$Mg &   $^{24}$Mg($\alpha$, $\gamma$)$^{28}$Si   &     -0.829  & -0.667 \\
$^{26}$Mg &   $^{26g}$Al($\alpha$, p)$^{29}$Si         &     -0.686  & -0.631 \\
          &   $^{26}$Si($\alpha$, p)$^{29}$P           &     -0.550  & -0.493 \\
\enddata
\end{deluxetable}

\begin{deluxetable}{lcc}
\small
\tablecaption{Same as Table 16, for Model F08. }
\tablehead{ 
\colhead{Isotope}   & 
\colhead{Reaction (Individual variations)}   & 
\colhead{Reaction (Monte Carlo)}    
}
\startdata
$^{12}$C &   $^{12}$C(p, $\gamma$)$^{13}$N    &   --        \\
         &   $^{12}$C($\alpha$, $\gamma$)$^{16}$O  &  --      \\
$^{16}$O  & $^{16}$O($\alpha$, $\gamma$)$^{20}$Ne  & $^{16}$O($\alpha$, $\gamma$)$^{20}$Ne      \\
      & $^{12}$C($\alpha$, $\gamma$)$^{16}$O    &    --    \\
$^{20}$Ne & $^{20}$Ne($\alpha$, $\gamma$)$^{24}$Mg & $^{20}$Ne($\alpha$, $\gamma$)$^{24}$Mg      \\
      & $^{12}$C($\alpha$, $\gamma$)$^{16}$O    &    --    \\
$^{24}$Mg & $^{24}$Mg($\alpha$, $\gamma$)$^{28}$Si & $^{24}$Mg($\alpha$, $\gamma$)$^{28}$Si     \\
      & $^{12}$C($\alpha$, $\gamma$)$^{16}$O    &   --     \\
\enddata
\end{deluxetable}

\begin{deluxetable}{lc}
\small
\tablecaption{Summary of the most influential nuclear processes, as collected from Tables 1--10. These reactions
              affect the yields of, at least, 3 isotopes 
              when their nominal rates are varied by a factor of 10 up and/or down. 
	       See text for details.}
\tablehead{ 
\colhead{Reaction} &  
\colhead{Models affected}    
}
\startdata
$^{12}$C($\alpha$, $\gamma$)$^{16}$O\tablenotemark{a}      & F08, K04-B2, K04-B4, K04-B5  \\
$^{18}$Ne($\alpha$, p)$^{21}$Na\tablenotemark{a}           & K04-B1\tablenotemark{b} \\
$^{25}$Si($\alpha$, p)$^{28}$P           & K04-B5 \\
$^{26g}$Al($\alpha$, p)$^{29}$Si         & F08  \\
$^{29}$S($\alpha$, p)$^{32}$Cl           & K04-B5 \\
$^{30}$P($\alpha$, p)$^{33}$S            & K04-B4 \\
$^{30}$S($\alpha$, p)$^{33}$Cl           & K04-B4\tablenotemark{b}, K04-B5\tablenotemark{b}  \\
$^{31}$Cl(p, $\gamma$)$^{32}$Ar          & K04-B1  \\
\enddata
\tablenotetext{a}{Reaction experimentally constrained to better than a factor of $\sim$10 at XRB temperatures. See Section 5.}
\tablenotetext{b}{Reaction that affects the total energy generation rate by more than 5\% at some time interval in this model, when its
                  rate is varied by a factor of 10 up and/or down. See text and Table 20 for details.}
\end{deluxetable}

\begin{deluxetable}{lc}
\small
\tablecaption{Nuclear processes affecting the total energy output
              by more than 5\%, as well as the yield of at least one isotope, when their nominal rates are 
	      individually varied by a factor of 10 up and/or down, for the given model. See text for details.}
\tablehead{ 
\colhead{Reaction} &  
\colhead{Models affected}    
}
\startdata
$^{15}$O($\alpha$, $\gamma$)$^{19}$Ne\tablenotemark{a}    &  K04, K04-B1, K04-B6 \\
$^{18}$Ne($\alpha$, p)$^{21}$Na\tablenotemark{a}           & K04-B1, K04-B6  \\
$^{22}$Mg($\alpha$, p)$^{25}$Al          & F08  \\
$^{23}$Al(p, $\gamma$)$^{24}$Si          & K04-B1  \\
$^{24}$Mg($\alpha$, p)$^{27}$Al\tablenotemark{a}          & K04-B2  \\
$^{26g}$Al(p, $\gamma$)$^{27}$Si\tablenotemark{a}          & F08  \\
$^{28}$Si($\alpha$, p)$^{31}$P\tablenotemark{a}           & K04-B4  \\
$^{30}$S($\alpha$, p)$^{33}$Cl           & K04-B4, K04-B5  \\
$^{31}$Cl(p, $\gamma$)$^{32}$Ar          & K04-B3  \\
$^{32}$S($\alpha$, p)$^{35}$Cl           & K04-B2  \\
$^{35}$Cl(p, $\gamma$)$^{36}$Ar\tablenotemark{a}          & K04-B2  \\
$^{56}$Ni($\alpha$, p)$^{59}$Cu          & S01  \\
$^{59}$Cu(p, $\gamma$)$^{60}$Zn          & S01  \\
$^{65}$As(p, $\gamma$)$^{66}$Se          & K04, K04-B2, K04-B3  \\
$^{69}$Br(p, $\gamma$)$^{70}$Kr          & S01  \\ 
$^{71}$Br(p, $\gamma$)$^{72}$Kr          & K04-B7 \\
$^{103}$Sn($\alpha$, p)$^{106}$Sb        & S01  \\
\enddata
\tablenotetext{a}{Reaction experimentally constrained to better than a factor of $\sim$10 at XRB temperatures. See Section 5.}
\end{deluxetable}

\begin{deluxetable}{lcccc}
\small
\tablecaption{Comparison between theoretical rates  from Tables 19 \& 20, according to different rate compilations.  
The table shows, for a particular reaction, the maximum disagreement between our rate and the 
REACLIBv0 rate over the temperature range covered in all our studies ($\sim$0.4 - 2.5 GK). All Q-values are given in keV. }  
\tablehead{ \colhead{Reaction} & \colhead{Q-value}   & \colhead{Q-value}    & 
              \colhead{Q-value} &  \colhead{Maximum} \\
                               & \colhead{Audi et al. 2003a} & \colhead{this work\tablenotemark{a}} &
	       \colhead{REACLIBv0\tablenotemark{a}} &  \colhead{disagreement}
 }
 \startdata
$^{22}$Mg($\alpha$, p)$^{25}$Al  & 3655(1)                    & 3655 [NS] & 3655 [NS] & 20\% \\
$^{25}$Si($\alpha$, p)$^{28}$P   & 6119(11)                   & 6122 [NS] & 6119 [NS] & 20\% \\
$^{26g}$Al($\alpha$, p)$^{29}$Si & 4820.68(6)                 & 4823 [NS] & 4820 [NS] &40\% \\
$^{29}$S($\alpha$, p)$^{32}$Cl   & 5306(50)                   & 5307 [NS] & 5306 [NS] & $\times$2 \\
$^{30}$S($\alpha$, p)$^{33}$Cl   & 2077(3)                    & 2076 [NS] & 2076 [NS] & $\times$2 \\
$^{30}$P($\alpha$, p)$^{33}$S    & 1521.36(34)                & 1522 [NS] & 1521 [NS] & 20\% \\
$^{31}$Cl(p, $\gamma$)$^{32}$Ar  & 2422(50)                   & 2404 [Her95]& 2404 [Her95] & 0 \\
$^{32}$S($\alpha$, $\gamma$)$^{36}$Ar& 6640.76(14)            & 6639 [NS] & 6641 [NS] & $\times$2 \\
\enddata
\tablenotetext{a}{Sources of the theoretical estimates: M=MOST Hauser-Feschbach code; NS=NON-SMOKER hauser-Feschbach code; 
                 For01=Forstner et al. (2001);
                 Fis04=Fisker's shell model calculation --see REACLIBv0 database; Her95=Herndl et al. (1995), 
		  shell model calculation.}
\tablenotetext{b}{Q-value estimated from systematics. See Audi et al. (2003a).}
\end{deluxetable}

\clearpage

 \begin{figure*}
 \centering
 \plotone{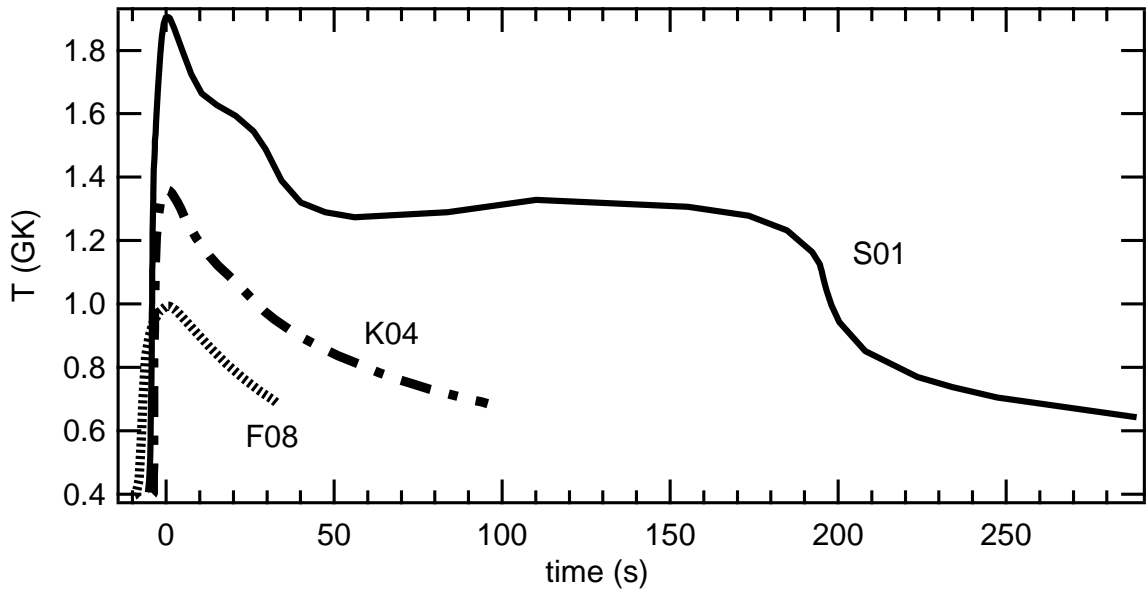}
 \caption{Temperature versus time profiles corresponding to models K04 (Koike et al. 2004), 
                 S01 (Schatz et al. 2001), and F08 (Fisker et al. 2008).}
 \label{Figure1} 
 \end{figure*}

\clearpage

 \begin{figure*}
 \centering
 \plotone{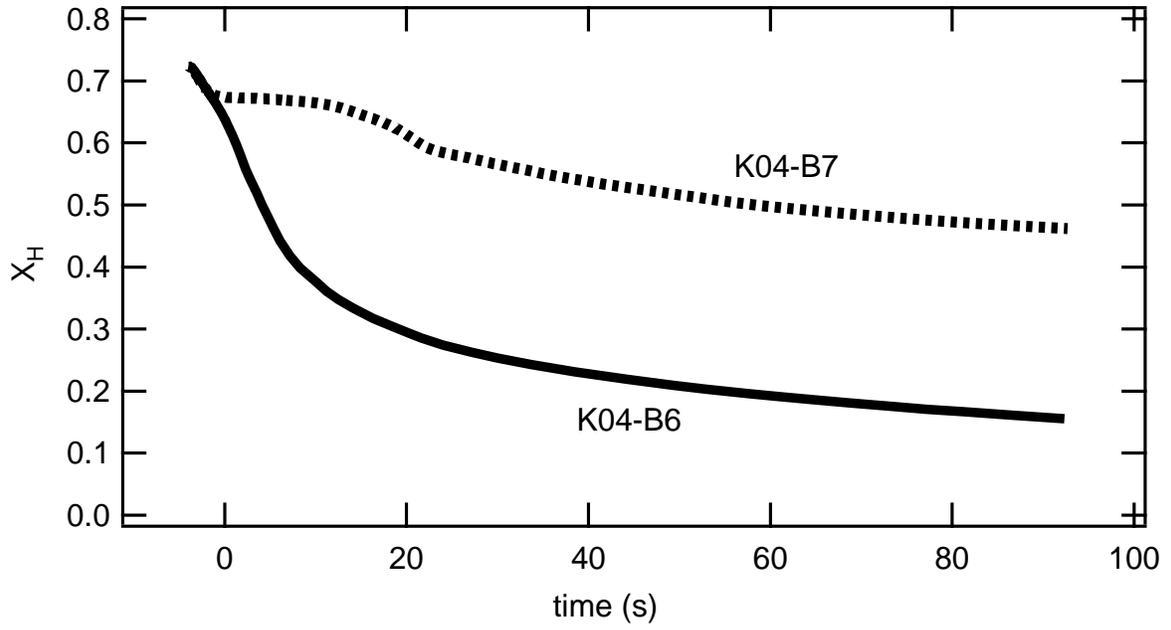}
 \caption{Time evolution of the hydrogen mass fraction in models K04-B6 (low T) and K04-B7 (high T).
          Models are based on Koike et al. (2004) (see text for details). t=0 corresponds to the time
	  at which $T_{peak}$ is reached.}
 \label{Figure2} 
 \end{figure*}

\clearpage

 \begin{figure*}
 \centering
 \plotone{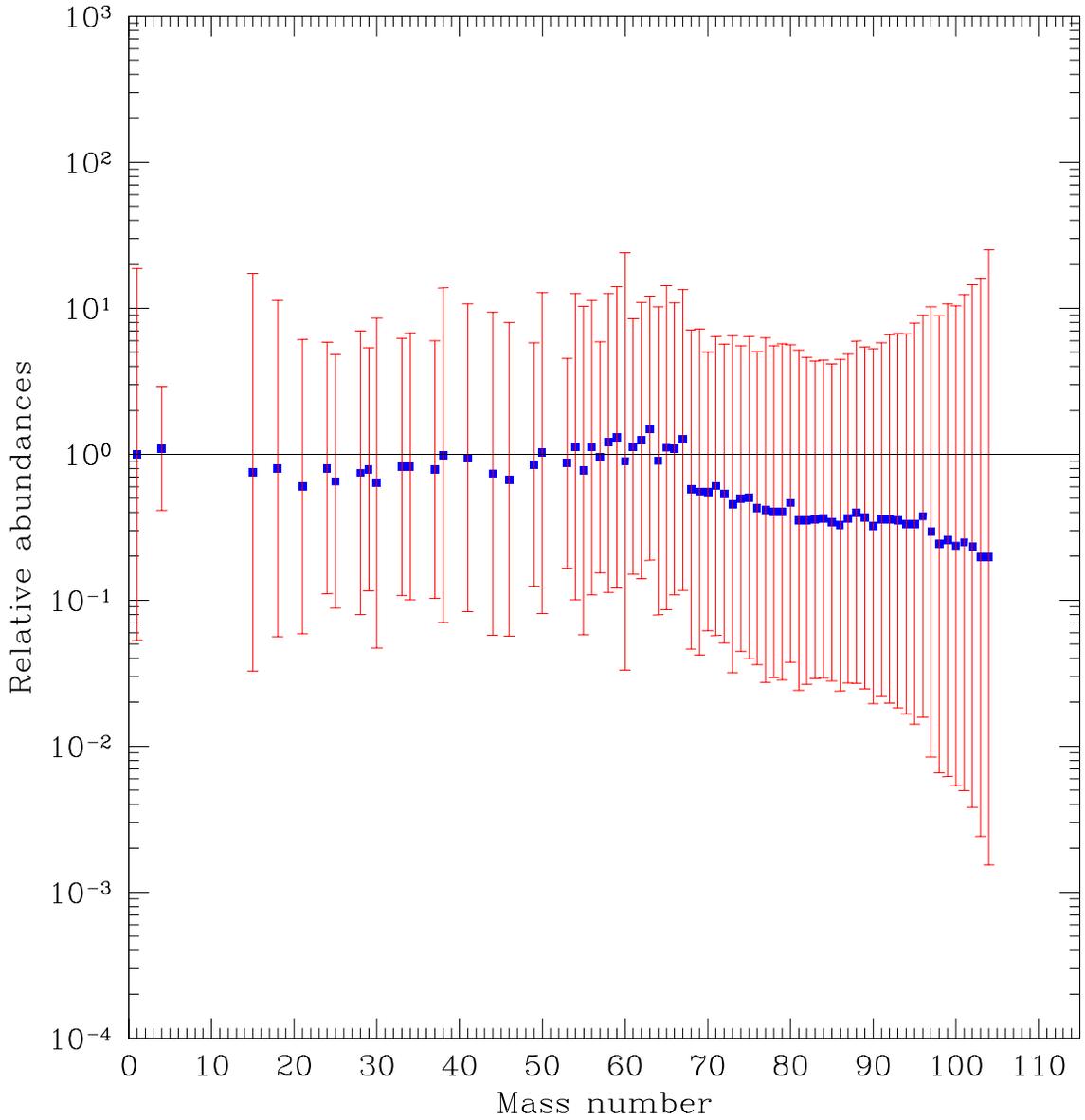}
 \caption{Uncertainties in the final distribution of abundances resulting from 
 the simultaneous variation of all nuclear processes in bulk by factors ranging roughly between 
 $0.1$ and $10$, for Model K04 (see Section 4.2 for details).}
 \label{Figure3} 
 \end{figure*}

\clearpage

 \begin{figure*}
 \centering
 \plotone{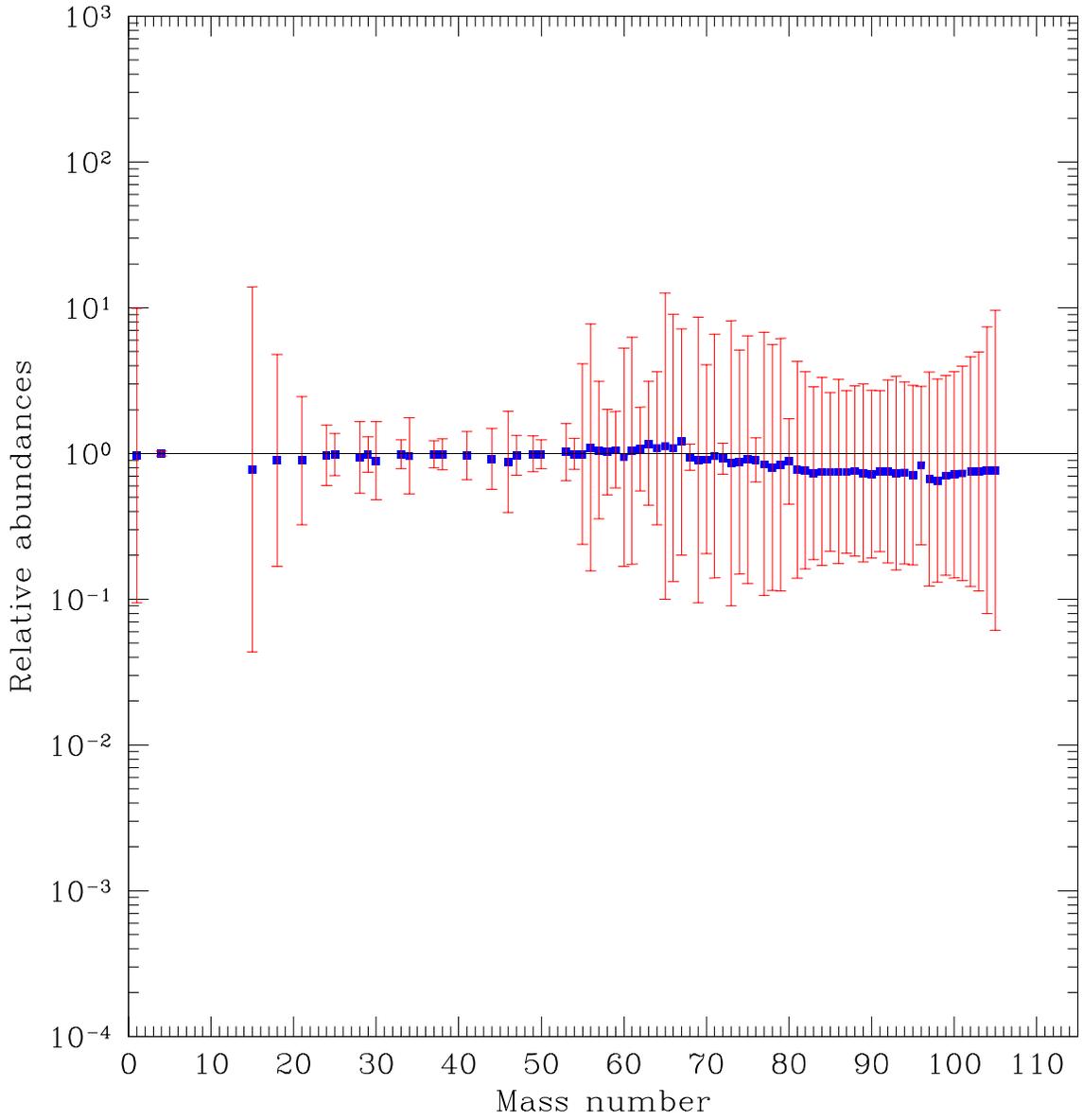}
 \caption{Same as Fig. 3, when all nuclear processes, except the 3$\alpha$ and all the $\beta$-decay rates, 
 are allowed to vary simultaneously by factors ranging roughly between $0.1$ and $10$, for Model K04 (see Section
 4.2 for details).}
 \label{Figure4} 
 \end{figure*}

\clearpage

 \begin{figure*}
 \centering
 \plotone{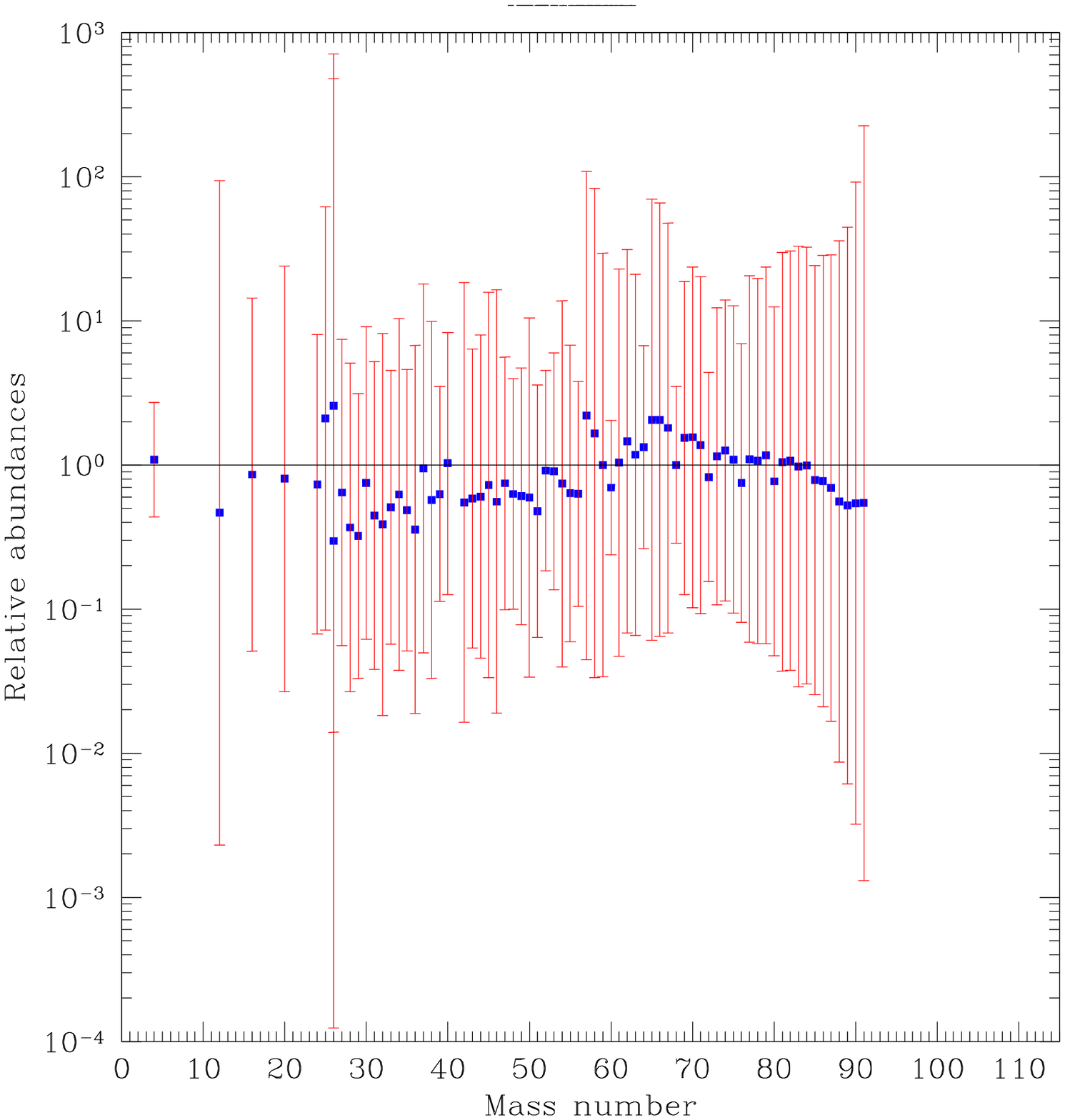}
 \caption{Same as Fig. 3, for Model F08.}
 \label{Figure5} 
 \end{figure*}

\clearpage

 \begin{figure*}
 \centering
 \plotone{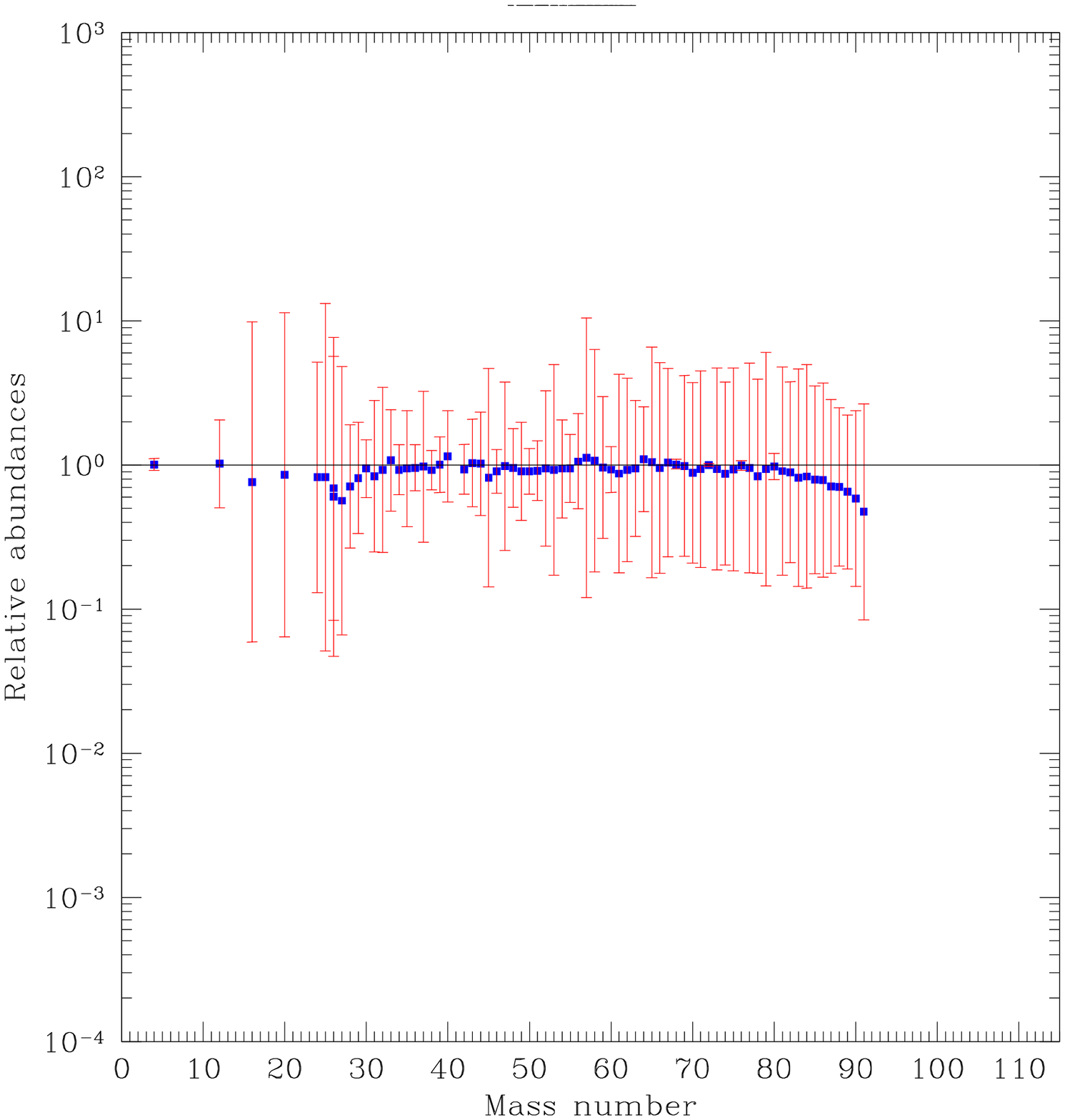}
 \caption{Same as Fig. 4, for Model F08.}
 \label{Figure6} 
 \end{figure*}

\clearpage

 \begin{figure*}
 \centering
 \plotone{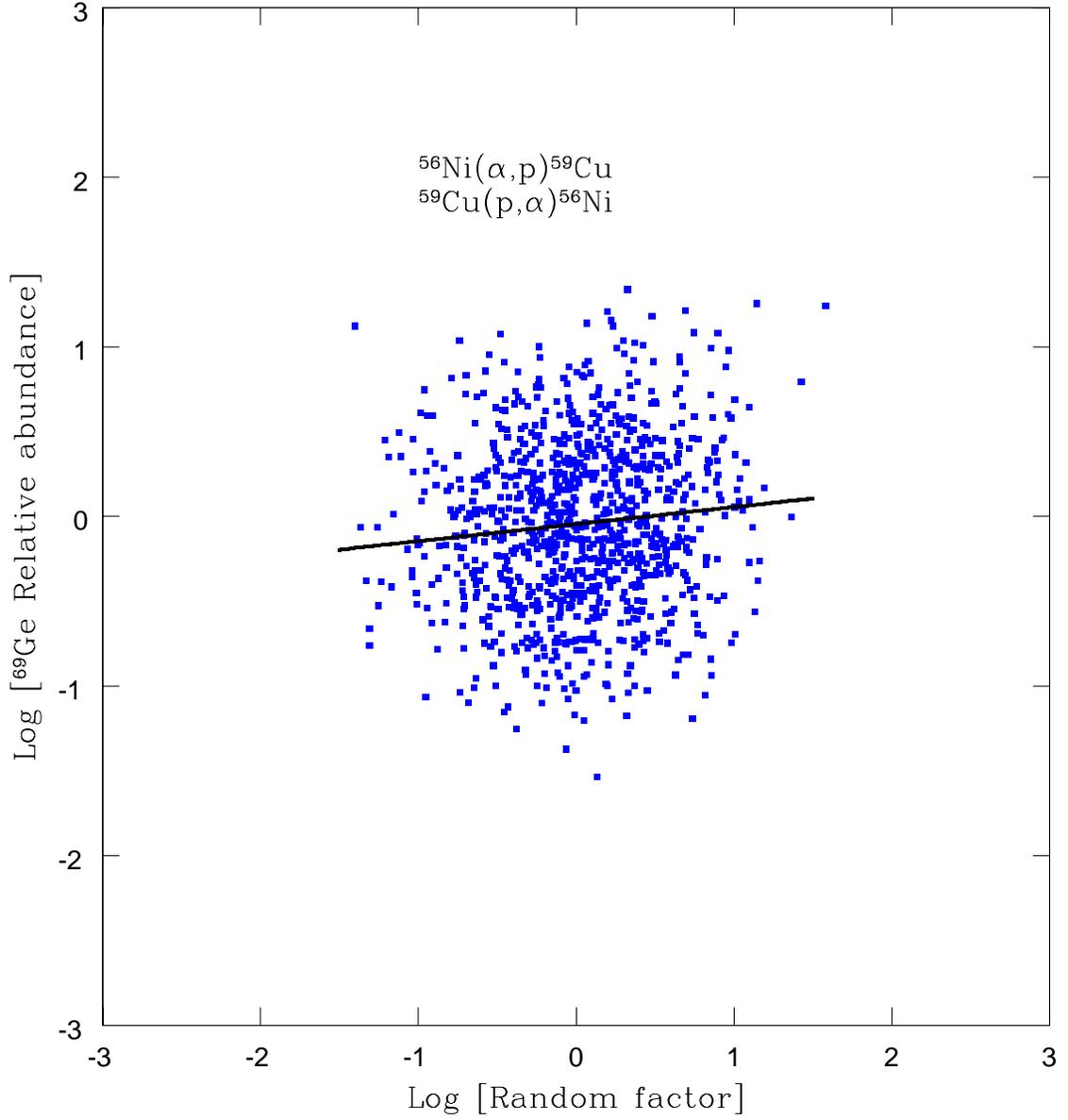}
 \caption{Correlation between the random enhancement factors applied to the $^{56}$Ni($\alpha$, p) rate
          and the final $^{69}$Ge normalized abundance, for Model K04.}
 \label{Figure7}
 \end{figure*}

 \clearpage

 \begin{figure*}
 \centering
 \plotone{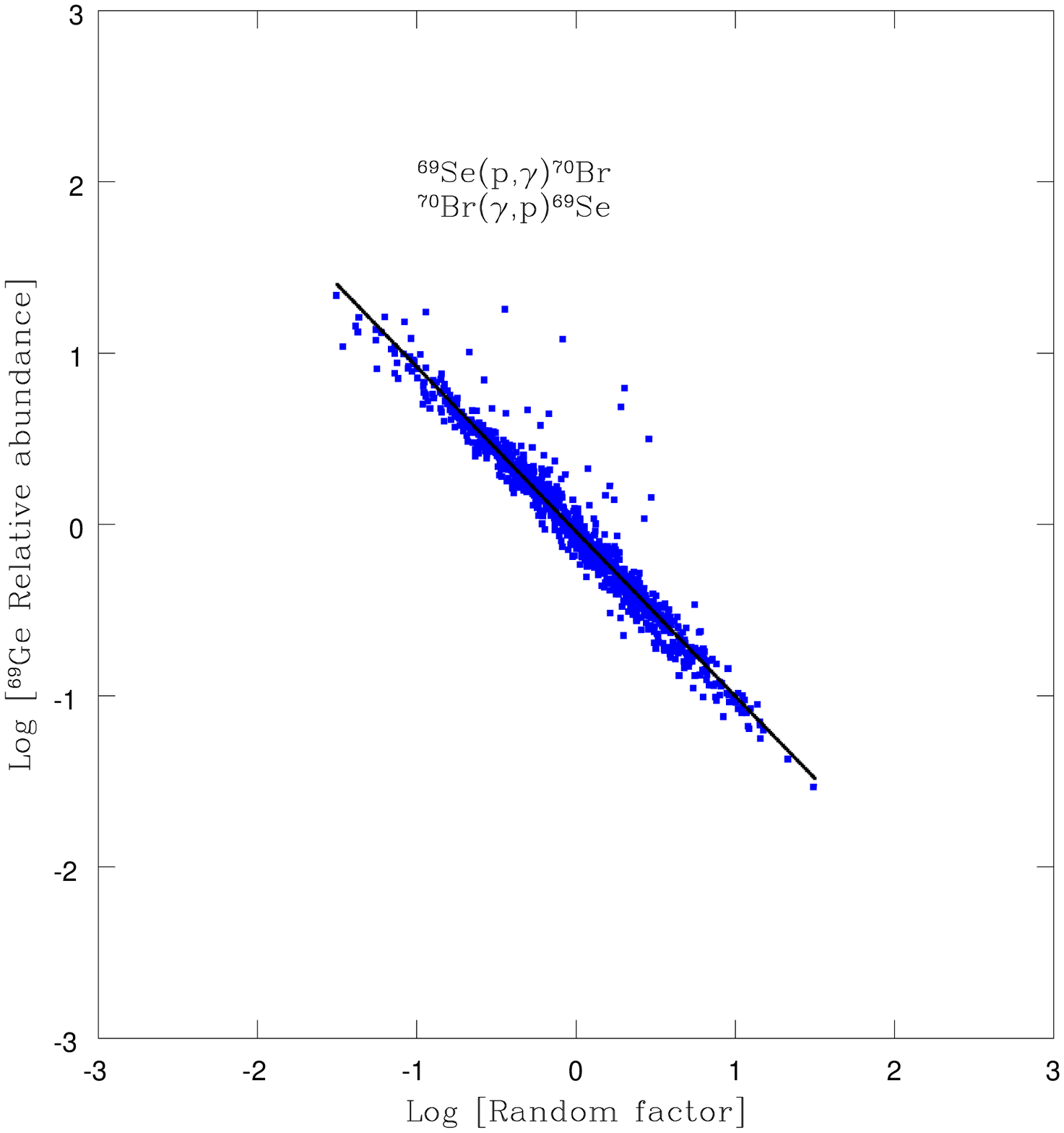}
 \caption{Same as Fig. 7, for the strong correlation between the $^{69}$Se(p, $\gamma$) rate and the final $^{69}$Ge abundance.}
 \label{Figure8}
 \end{figure*}

 \clearpage

 \begin{figure*}
 \centering
 \plotone{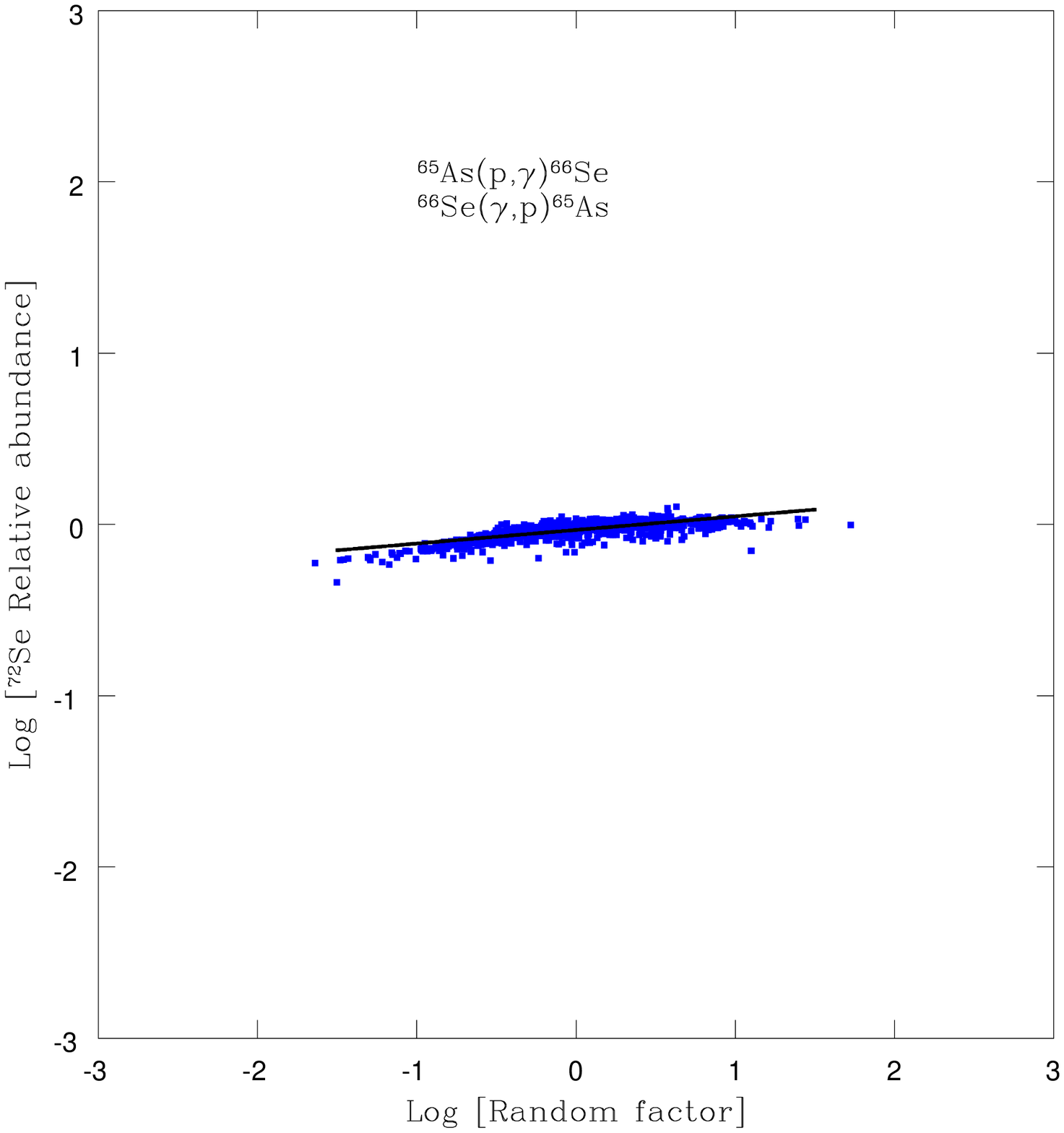}
 \caption{Same as Fig. 7, for the strong correlation between the $^{65}$As(p, $\gamma$) rate and the final $^{72}$Se abundance.}
 \label{Figure9}
\end{figure*}

\end{document}